\documentclass[fleqn,usenatbib]{mnras}

\DeclareRobustCommand{\Yang}[3]{#2}
\let\Yangthebibliography\thebibliography
\def\thebibliography{\DeclareRobustCommand{\Yang}[3]{##3}\Yangthebibliography}

\usepackage[T1]{fontenc}
\usepackage{newtxtext,newtxmath}
\usepackage{url}
\usepackage[english]{babel}
\usepackage{ulem}
\usepackage{caption}
\usepackage{xcolor}
\usepackage{multirow}
\usepackage{adjustbox}
\usepackage{booktabs}
\usepackage{gensymb}
\usepackage{graphicx}	
\usepackage{amsmath}	
\usepackage{array}
\PassOptionsToPackage{colorlinks=true, allcolors=blue}{hyperref}
\usepackage{hyperref}

\title[Measuring the Spin of 4U 1543-47]{Measuring the Spin of Black Hole Transient 4U 1543-47 Using \textit{Insight}-HXMT}

\author[Jun Yang et al.]{
Jun Yang$^{1,2}$,
Nan Jia$^{1,2}$,
Erlin Qiao$^{1,2}$,
Yujia Song$^{1,2}$,
Lijun Gou$^{1,2}$\thanks{E-mail: lgou@nao.cas.cn}
\\
$^{1}$Key Laboratory for Computational Astrophysics, National Astronomical Observatories, Chinese Academy of Sciences, Datun Road A20, Beĳing 100012, China\\
$^{2}$School of Astronomy and Space Sciences, University of Chinese Academy of Sciences, Datun Road A20, Beĳing 100049, China\\
}

\date{Accepted XXX. Received YYY; in original form ZZZ}

\pubyear{2015}

\begin{document}
\label{firstpage}
\pagerange{\pageref{firstpage}--\pageref{lastpage}}
\maketitle

\begin{abstract}
We provided a comprehensive study of the properties of the black hole in the low-mass X-ray binary system 4U 1543-47, specifically focusing on the 2021 outburst (MJD 59380-59470). Using observations from the \textit{Insight}-HXMT mission, we employed X-ray reflection fitting method and analyzed spectral data to estimate key black hole parameters. Through our investigation redbased on 6 out of the 52 available observations, we estimated the spin parameter of the black hole to be $0.902_{-0.053}^{+0.054}$ and the inclination angle of the accretion disk to be $28.91_{-1.24}^{+1.82}$ degrees (90\% confidence limits, statistical only), then we discussed the influence of high luminosity. Based on the \texttt{relxill} series models are not suitable for thick disk scenario, and in comparison with findings from other studies, we propose that our estimation of the spin value may be exaggerated.
\end{abstract}




\section{INTRODUCTION}
 
Black holes, predicted by the theory of general relativity,  are regions in space characterized by an event horizon, beyond which nothing can escape. The event horizon represents the boundary around a black hole, beyond which the escape velocity exceeds the speed of light. Using the Kerr metric \citep{PhysRevLett.11.237}, the complex nature of black holes and the surrounding space-time can be accurately described in terms of mass and spin. For a black hole with mass $M$ and angular momentum $J$, the dimensionless spin parameter $a_*$ is defined as $cJ/GM^2$, where $c$ is the speed of light and $G$ is the gravitational constant. This parameter provides a measure of how fast the black hole is spinning.

The innermost stable circular orbit (ISCO) is a critical characteristic of a black hole, fundamentally influencing the physics of black hole accretion. The ISCO is defined by the black hole's intrinsic properties—specifically its mass and spin \citep{1972ApJ...178..347B,2019NatAs...3...41R}—and represents the closest orbit in which matter can stably orbit the black hole. As accreted matter goes across the ISCO, it is inevitably pulled towards the event horizon, establishing the ISCO as the maximum inner boundary of the accretion disk. Consequently, the spin of the black hole can be inferred from the monotonic relationship between the inner radius of the disk and the ISCO. Two prevalent methods for measuring the black hole's spin are X-ray reflection fitting \citep{1989MNRAS.238..729F,2002ApJ...570L..69M} and thermal continuum fitting \citep{1997ApJ...482L.155Z}.

X-ray reflection fitting relies on observing gravitational redshifts in atomic features within the X-ray spectrum. Conversely, thermal continuum fitting uses the temperature profile of the accretion disk to infer the location of the ISCO. By performing spectroscopic measurements to determine the inner radius of the accretion disk, the ISCO can be identified, thereby enabling an estimation of the black hole's spin.

4U 1543-47 is a transient low-mass X-ray binary system comprising a black hole and an A2V companion star \citep{2006MNRAS.371.1334R}. \citet{2004ApJ...610..378P} reported that the black hole has a mass of \(9.4 \pm 1.0 \ \rm{M_\odot}\) and is located \(7.5 \pm 1.0\) kpc from Earth, with a low orbital inclination of \(20.7 \pm 1.5\) degrees. The Uhuru satellite first identified this system during an outburst in 1971 \citep{1972ApJ...174L..53M}. Since its initial discovery, the system has exhibited a pattern of outbursts approximately every decade, with recorded eruptions in 1983 \citep{1984PASJ...36..799K}, 1992 \citep{1992IAUC.5504....1H}, and 2002 \citep{2004ApJ...610..378P}, although no significant activity was noted around 2012. This study concentrates on the most recent outburst, which occurred in June 2021.

The target source 4U 1543-47 has been extensively studied, primarily using \textit{MAXI} data from the 2002 burst. \citet{2006ApJ...636L.113S} were the first to use the continuum-fitting method and estimated the spin value to be \(0.80^{+0.05}_{-0.05}\), assuming the inclination angle of the accretion disk to be equal to the binary orbital inclination angle of \(20.7 \pm 1.5\) degrees. Using the same inclination, \citet{2014ApJ...793L..33M} reported a spin measurement of \(0.43^{+0.22}_{-0.31}\). \citet{2009ApJ...697..900M} obtained a spin of \(0.3 \pm 0.1\) with an inclination of \(32^{+3}_{-4}\) degrees, constrained by their fits to the iron line, using a combination of the continuum-fitting and X-ray reflection fitting methods. Research by \citet{2020MNRAS.493.4409D} investigated the outburst using the X-ray reflection method, revealing a spin value of \(0.67^{+0.15}_{-0.08}\) and measuring the disk inclination to be \(36.3^{+5.3}_{-3.4}\) degrees with data from the \textit{RXTE} mission. \citet{2023ApJ...946...19D} investigated the 2021 outburst with \textit{NuSTAR} observations, using the X-ray reflection method, and measured a spin of \(0.98^{+0.01}_{-0.02}\) and an inclination angle of \(68^{+3}_{-4}\) degrees. Additionally, \citet{2023A&A...677A..79Y} measured the spin value to be \(0.65^{+0.14}_{-0.24}\), and \citet{2023MNRAS.tmp.2981C} reported a spin value of \(0.42 \pm 0.12\), both employing the continuum-fitting method.


The outburst of the target source 4U 1543-47 was detected by \textit{MAXI} on June 11, 2021 \citep{2021ATel14701....1N}, and it quickly reached the soft state. According to the results of \citet{2004ApJ...610..378P}, assuming the mass of the black hole is \(9.4 \ \rm{M_\odot}\) and the distance is 7.5 kpc, its Eddington luminosity is approximately \(1.18 \times 10^{39} \ \rm{erg/s}\). \citet{2023A&A...673A.104S} conducted a comprehensive analysis of the optical-infrared (OIR) spectroscopy of 4U 1543–47 during this outburst, providing unique insights into the OIR spectra of black hole transients in their ultra-luminous state. In subsequent operations, it is essential to compare and avoid excessive brightness to ensure the reliability of data verification, as excessive brightness could potentially compromise the results.

The paper is organized as follows. In Section 2, we present our data reduction and selection procedures. In Section 3, we discuss our spectral analysis. In Section 4 we present our discussions and describe the spin result. Lastly, in Section 5 we summarize our conclusions.

\section{OBSERVATIONS AND DATA REDUCTION}

\subsection{Observations}

The \textit{Hard X-ray Modulation Telescope} (\textit{Insight}-HXMT) satellite is China's first space astronomy satellite \citep{2020SCPMA..6349502Z}, launched in June 2017. It carries three telescopes: the Low Energy X-ray Telescope (LE, 1-15 keV) \citep{2020SCPMA..6349505C}, the Medium Energy X-ray Telescope (ME, 5-30 keV) \citep{2020SCPMA..6349504C}, and the High Energy X-ray Telescope (HE, 20-250 keV) \citep{2020SCPMA..6349503L}. The \textit{Insight}-HXMT mission, with its large effective area, broad energy range, and satisfactory energy resolution, provides critical capabilities for studying X-ray black hole binaries. These characteristics enable precise measurements and in-depth studies of various features, including the spin of black holes \citep{2020JHEAp..27...53Z,2021ApJ...916..108Z,2022MNRAS.512.2082L}.

The \textit{Insight}-HXMT satellite observed 4U 1543-47 from June 15, 2021 (MJD 59380) to September 19, 2021 (MJD 59470), comprising a total of 52 observations.

\subsection{Data Reduction}

In the preliminary processing of the data, we followed the HXMT manual\footnote{\url{http://hxmtweb.ihep.ac.cn/SoftDoc.jhtml}} and extracted data using the pipeline provided by the HXMTDAS V2.05 software\footnote{\url{http://hxmtweb.ihep.ac.cn/software.jhtml}}. We filtered the spectrum using the following good time interval (GTI) settings: (1) the elevation angle (ELV) is greater than \(10\degree\); (2) the ELV above the bright Earth for the LE detector is greater than \(30\degree\); (3) the value of the geomagnetic cutoff rigidity (COR) is greater than 8 GeV; (4) the offset angle from the pointing direction is less than \(0.04\degree\); and (5) the time before and after the South Atlantic Anomaly passage is greater than 300 seconds.

Firstly, we used the "addspec" method provided by FTOOLS\footnote{\url{https://heasarc.gsfc.nasa.gov/ftools/}} to merge the spectra that have two or more exposure IDs. Subsequently, we generated a new response file using the "addrmf" method, also from FTOOLS. During the spectral data processing, we initially utilized "grppha" for channel rebinning, ensuring that each bin possesses a minimum of 100 counts.

\begin{figure}
    \centering
    \includegraphics[width=1\linewidth]{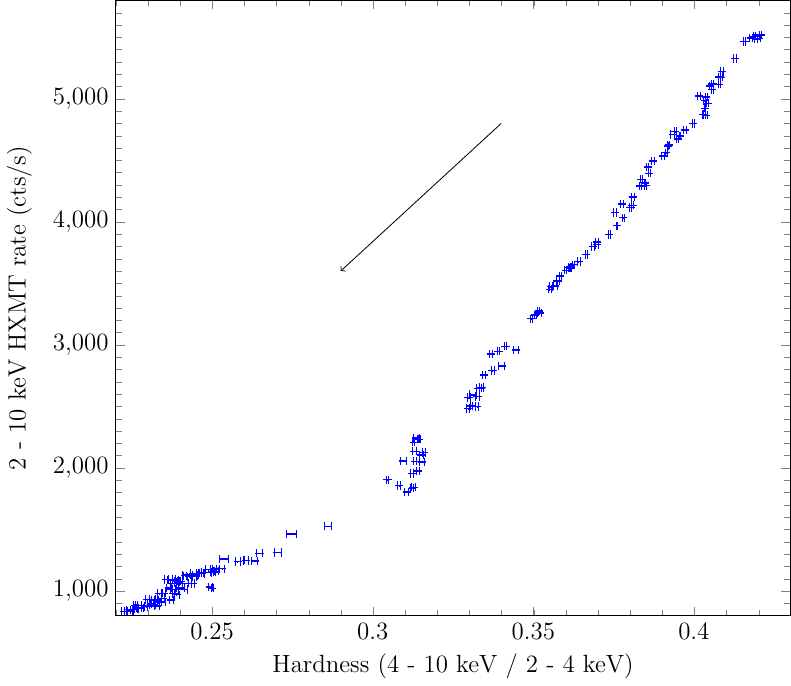}
    \caption{Hardness-intensity diagram of the outburst. The hardness ratio between the 2 - 4 keV and 4 - 10 keV bands is indicated on the horizontal axis, while the photon count rate within the 2 - 10 keV energy band is on the vertical axis. As depicted in the figure, the trajectory follows a pattern from the top right to the bottom left, indicating that the source is becoming progressively softer, leading to a reduction of the reflection feature.}
  \label{fig:1}
\end{figure}

The evolution of hardness is depicted in Figure \ref{fig:1}, demonstrating a trajectory from the top right to the bottom left. This figure indicates that the source initiates in a soft state and progressively becomes softer as the flux weakens. Due to its high luminosity, the initial conditions may undergo drastic changes. To avoid potential inconsistencies, we attempt to sidestep the use of initial data for studying the black hole spin.

Moreover, as the source progressively becomes softer, the reflection feature weakens. Consequently, we consciously avoid including observations from the latter stages. After careful selection, the observations utilized in this analysis are presented in Table \ref{tab:1}.

\begin{table*}
\caption{Summary of the \textit{Insight}-HXMT observations of the black hole X-ray binary system 4U 1543-47 utilized in this study. We have estimated the flux in the 1 - 10 keV energy band using the model \texttt{const*tbabs*(diskbb+powerlaw)*gabs}. The Eddington luminosity is calculated by assuming the mass of the black hole is $9.4$ $ \rm{M_\odot}$, and the distance is $7.5$ kpc.}
\label{tab:1}
\begin{tabular}{rrcccccc}
\toprule
  \multirow{2}{*}{Spec.} &  \multirow{2}{*}{ObsID} &   \multirow{2}{*}{MJD}    &  Count rate &  Count rate &  Hardness & Flux & L\\
 & & & (cts/s 2-4keV) & (cts/s 4-10keV) & (4-10/2-4) & ($\times10^{-7}\;\rm{erg\;cm^{-2}\;s^{-1}}$) & ($L_{\rm{Edd}}$) \\
\midrule
1  &  P0304155001 &  59385.09 &    2961 &     1134 &  0.383 & 21.839 & 1.246 \\
2  &  P0304155002 &  59387.08 &    2539 &      909 &  0.358 & 18.458 & 1.053 \\
3  &  P0304155003 &  59389.86 &    2211 &      760 &  0.344 & 16.112 & 0.919 \\
4  &  P0304026005 &  59390.06 &    2184 &      743 &  0.340 & 15.935 & 0.909 \\
5  &  P0304026006 &  59391.25 &    2068 &      700 &  0.339 & 15.112 & 0.862 \\
6  &  P0304155004 &  59392.31 &    1976 &      661 &  0.335 & 14.423 & 0.823 \\
7  &  P0304026007 &  59392.51 &    1975 &      662 &  0.335 & 14.447 & 0.824 \\
8  &  P0304026008 &  59393.11 &    1924 &      640 &  0.333 & 13.869 & 0.791 \\
9  &  P0304155005 &  59393.90 &    1868 &      622 &  0.333 & 13.659 & 0.779 \\
10 &  P0304026009 &  59394.10 &    1855 &      615 &  0.332 & 13.443 & 0.767 \\
11 &  P0304026010 &  59395.56 &    1684 &      530 &  0.315 & 12.233 & 0.698 \\
12 &  P0304155006 &  59396.95 &    1605 &      507 &  0.316 & 11.679 & 0.666 \\
13 &  P0304026011 &  59397.94 &    1554 &      488 &  0.314 & 11.371 & 0.649 \\
14 &  P0304155007 &  59399.27 &    1488 &      468 &  0.315 & 10.971 & 0.626 \\
15 &  P0304026012 &  59400.26 &    1424 &      439 &  0.308 & 10.409 & 0.594 \\
\bottomrule
\end{tabular}
\end{table*}

\section{SPECTRAL ANALYSIS}

\begin{figure}
    \centering
    \includegraphics[width=1\linewidth]{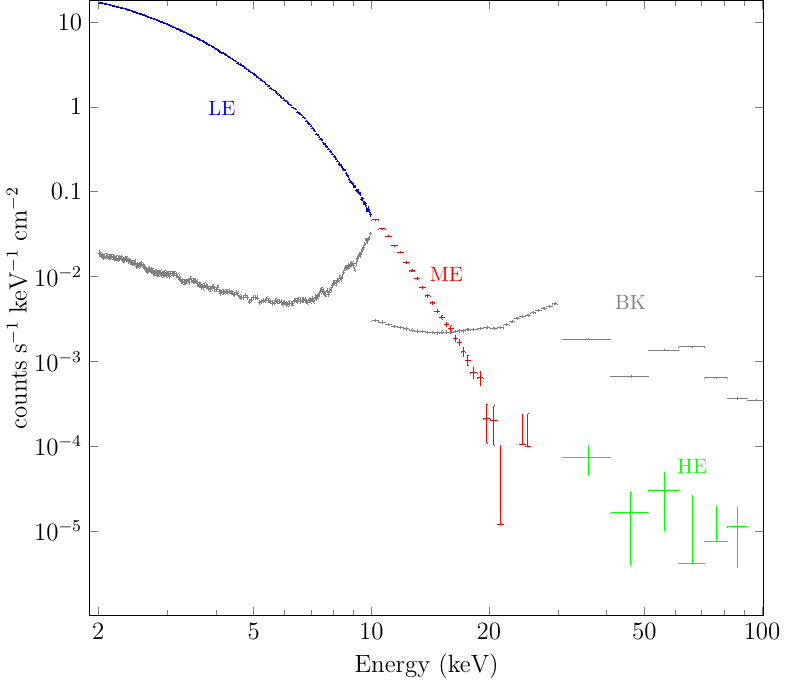}
     \caption{The spectrum and background of Spectrum 1, illustrated on a logarithmic y-axis, showing the source's current position in a soft state. While the LE spectrum data is significantly strong, the HE data in this state is markedly weak—so much so that it falls below the background level, leading to its exclusion from our analysis.}
  \label{fig:2}
\end{figure}

As displayed in Figure \ref{fig:2}, we plotted the observation data of P304155001. When contrasted with the case study in \citet{2006ARA&A..44...49R}, it becomes apparent that the source is in the soft state. It's worth noting that the photon count in the HE observation data is relatively scarce. 
In contrast to the 0.1 photons at 10 keV observed by LE, the HE detection is significantly lower, amounting to less than \(10^{-4}\) photons at 30 keV. This implies that both the energy and photon number distributions of the burst mainly occur below 10 keV. This figure also includes the background spectra, which visibly drowns out the HE data. Consequently, we decided to omit HE observations, focusing on these energy bands: 2 - 10 keV for LE and 10 - 30 keV for ME.

\subsection{The non-relativistic spectral model}

We employed XSPEC v12.12.0 \citep{1996ASPC..101...17A} to conduct the spectrum fitting. We started with a phenomenological model, \texttt{const*tbabs*(diskbb+powerlaw)}. Here, \texttt{const} is an energy-independent factor used to calibrate the difference between LE and ME. \texttt{tbabs} is an interstellar medium (ISM) absorption model \citep{2000ApJ...542..914W}, with the hydrogen column density parameter fixed at \(0.4 \times 10^{22} \, {\rm cm}^{-2}\) based on \citet{2004ApJ...610..378P} and \citet{2014ApJ...793L..33M}. \texttt{diskbb} is used to describe the accretion disk consisting of multiple blackbody components \citep{1984PASJ...36..741M}.

\begin{figure}
    \includegraphics[width=0.95\linewidth]{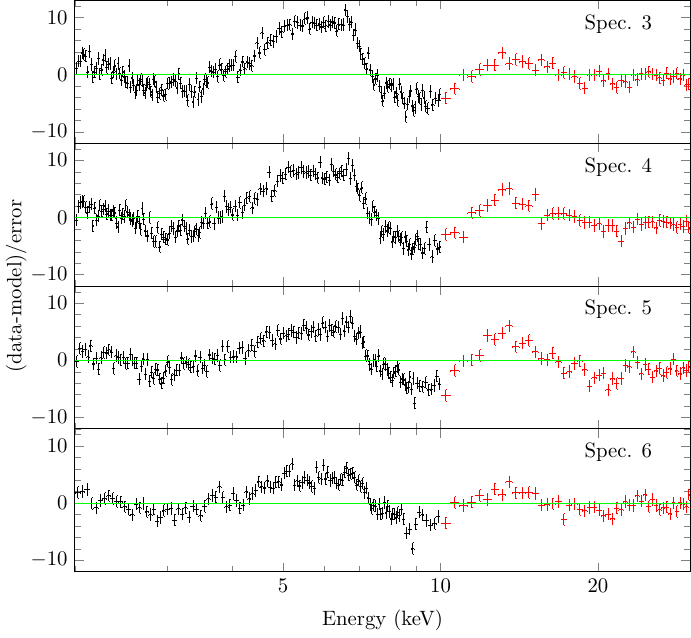}
    \caption{Residual plots of Spectrum 3 to 6, created by initially ignoring the 5 - 8 keV range and reintroducing it after fitting the spectra with the model \texttt{const*tbabs*(diskbb+powerlaw)}. The plots reveal a distinct iron line structure, suggesting that we can proceed with an analysis of these spectral reflection features. Additionally, strong absorption features are evident in the 8 to 11 keV range.}
    \label{fig:3}
\end{figure}

To investigate the existence of iron line components, we initially excluded the spectral energy range of 5 - 8 keV prior to fitting the model. Once the fitting was complete, we reintegrated this energy band into the total spectrum. This method enhances the signals indicative of an iron line presence, enabling a more intuitive observation of a potential reflection spectrum component. The $\chi^2/\nu$ values are presented in Table \ref{tab:3}. Subsequently, we observed the presence of a reflection spectrum iron line, as demonstrated in several instances within Figure \ref{fig:3}.

In the meantime, we found a prominent absorption spectrum within the 8 to 11 keV range over an extended duration. This observation necessitated the addition of a Gaussian absorption line model, denoted as \texttt{gabs}. This model comprises three parameters: the line energy in keV ($\rm{Line_E}$), line width in keV ($\sigma$), and line depth (Strength). A detailed analysis of this component was carried out by \citet{2023MNRAS.520.4889P}, who investigated several possible sources for the absorption feature observed in the spectrum. The study inferred that highly ionized and rapidly moving disk winds are the likely contributors to this absorption.

Incorporating \texttt{gabs} resulted in our model transforming into \texttt{const*tbabs*(diskbb+powerlaw)*gabs}.
Building upon the previous model, we retained the same parameters and added \texttt{gabs} to fit the absorption features. 
The results of the best fit are presented in Table \ref{tab:2}. This table shows that as the spectrum evolves, the temperature progressively decreases while the powerlaw index rises, indicating a softening of the spectrum. The optical depth at the line center in model \texttt{gabs} is calculated as $\rm{Strength}/\sigma / \sqrt{2\pi}$, which also exhibits a gradual increase over time.

We used this model to compute the spectral luminosity. The "flux" method in XSPEC was employed to calculate the flux in the 1 to 10 keV range. Assuming a distance of 7.5 kpc to the source, we estimated its luminosity. The results, along with the ratio to the Eddington limit, are presented in Table \ref{tab:1}.

\begin{table*}
\caption{This table shows the best-fitting parameters for the model \texttt{const*tbabs*(diskbb+powerlaw)*gabs}. The errors provided represent the 90\% confidence interval. For Spectrum 15, error calculation using the "error" method is not possible due to $\chi^2/\nu > 2$.}
\label{tab:2}
\begin{adjustbox}{width=480pt}
\begin{tabular}{c|cc|cc|ccc|c|ccc}
\toprule

 \multirow{2}{*}{Spec.} & \multicolumn{2}{c|}{\texttt{diskbb}} & \multicolumn{2}{c|}{\texttt{powerlaw}}  & \multicolumn{3}{c|}{\texttt{gabs}}  &  \multirow{2}{*}{\texttt{const}} & \multirow{2}{*}{$\chi^2$} &  \multirow{2}{*}{$\nu$}  & \multirow{2}{*}{$\chi^2/\nu$}   \\

 &     $T_{\rm in}$ (keV) &                     $N_{\rm bb}$  &               $\Gamma$  &                 $N_{\Gamma}$  &                   $\rm{Line_E}$ (keV) &                  $\sigma$ (keV) &               Strength &                  &    &   &   \\
\midrule
 1 & $1.207_{-0.002}^{+0.002}$ & $6373.24_{-49.84}^{+49.99}$ & $3.73_{-0.04}^{+0.04}$ & $37.70_{-1.84}^{+1.87}$ &  $9.55_{-0.14}^{+0.15}$ & $1.11_{-0.08}^{+0.08}$ & $0.70_{-0.08}^{+0.09}$ & $0.88_{-0.02}^{+0.02}$ &      1267.17 & 1173 &  1.080 \\
 2 & $1.155_{-0.002}^{+0.002}$ & $6463.63_{-67.60}^{+66.89}$ & $3.50_{-0.03}^{+0.03}$ & $30.50_{-1.69}^{+1.71}$ &  $9.61_{-0.17}^{+0.16}$ & $1.20_{-0.09}^{+0.09}$ & $0.99_{-0.12}^{+0.12}$ & $0.92_{-0.03}^{+0.03}$ &      1361.97 & 1247 &  1.094 \\
 3 & $1.133_{-0.002}^{+0.002}$ & $6031.21_{-68.42}^{+68.09}$ & $3.54_{-0.02}^{+0.02}$ & $31.15_{-1.53}^{+1.53}$ & $10.07_{-0.14}^{+0.12}$ & $1.51_{-0.07}^{+0.07}$ & $1.66_{-0.16}^{+0.15}$ & $0.99_{-0.03}^{+0.02}$ &      1480.59 & 1249 &  1.185 \\
 4 & $1.126_{-0.002}^{+0.002}$ & $6244.81_{-63.82}^{+63.11}$ & $3.82_{-0.10}^{+0.11}$ & $32.69_{-2.77}^{+3.01}$ &  $9.88_{-0.18}^{+0.23}$ & $1.34_{-0.08}^{+0.10}$ & $1.46_{-0.19}^{+0.27}$ & $0.97_{-0.04}^{+0.06}$ &      1320.37 & 1240 &  1.065 \\
 5 & $1.121_{-0.002}^{+0.002}$ & $6094.84_{-54.74}^{+54.55}$ & $4.20_{-0.05}^{+0.05}$ & $35.60_{-2.23}^{+2.28}$ &  $9.62_{-0.15}^{+0.14}$ & $1.17_{-0.08}^{+0.08}$ & $0.99_{-0.12}^{+0.12}$ & $0.96_{-0.02}^{+0.02}$ &      1478.89 & 1220 &  1.212 \\
 6 & $1.115_{-0.003}^{+0.003}$ & $5972.27_{-74.10}^{+73.86}$ & $4.21_{-0.07}^{+0.07}$ & $34.44_{-3.04}^{+3.07}$ &  $9.62_{-0.38}^{+0.25}$ & $1.20_{-0.19}^{+0.13}$ & $1.01_{-0.26}^{+0.22}$ & $0.97_{-0.05}^{+0.04}$ &      1193.74 & 1173 &  1.018 \\
 7 & $1.115_{-0.001}^{+0.001}$ & $5988.64_{-42.19}^{+42.39}$ & $4.26_{-0.06}^{+0.06}$ & $34.68_{-1.92}^{+2.01}$ &  $9.34_{-0.14}^{+0.16}$ & $1.07_{-0.08}^{+0.08}$ & $0.78_{-0.09}^{+0.10}$ & $0.91_{-0.02}^{+0.02}$ &      1433.54 & 1251 &  1.146 \\
 8 & $1.109_{-0.003}^{+0.003}$ & $5883.22_{-81.74}^{+78.30}$ & $3.81_{-0.04}^{+0.04}$ & $26.05_{-1.97}^{+2.04}$ & $10.25_{-0.19}^{+0.17}$ & $1.54_{-0.10}^{+0.11}$ & $1.59_{-0.21}^{+0.21}$ & $1.05_{-0.03}^{+0.03}$ &      1603.43 & 1196 &  1.341 \\
 9 & $1.113_{-0.002}^{+0.002}$ & $5696.84_{-50.48}^{+50.96}$ & $4.25_{-0.06}^{+0.06}$ & $33.54_{-2.09}^{+2.10}$ &  $9.74_{-0.27}^{+0.20}$ & $1.29_{-0.12}^{+0.09}$ & $1.38_{-0.25}^{+0.22}$ & $0.98_{-0.05}^{+0.04}$ &      1414.09 & 1227 &  1.152 \\
 10 & $1.105_{-0.003}^{+0.003}$ & $5872.74_{-97.73}^{+96.74}$ & $4.16_{-0.09}^{+0.09}$ & $27.80_{-3.49}^{+3.69}$ &  $9.53_{-0.31}^{+0.28}$ & $1.18_{-0.16}^{+0.15}$ & $0.98_{-0.22}^{+0.24}$ & $0.90_{-0.04}^{+0.04}$ &      1056.27 & 1103 &  0.958 \\
 11 & $1.082_{-0.002}^{+0.002}$ & $5576.77_{-47.76}^{+49.76}$ & $3.80_{-0.03}^{+0.03}$ & $30.21_{-1.22}^{+1.16}$ &  $9.66_{-0.19}^{+0.16}$ & $1.36_{-0.09}^{+0.07}$ & $1.74_{-0.21}^{+0.19}$ & $0.95_{-0.04}^{+0.04}$ &      1951.12 & 1277 &  1.528 \\
 12 & $1.086_{-0.003}^{+0.003}$ & $5292.49_{-68.80}^{+69.17}$ & $3.98_{-0.06}^{+0.06}$ & $30.90_{-2.02}^{+2.07}$ &  $9.25_{-0.15}^{+0.19}$ & $1.20_{-0.08}^{+0.09}$ & $1.60_{-0.17}^{+0.23}$ & $0.90_{-0.04}^{+0.05}$ &      1626.02 & 1099 &  1.480 \\
 13 & $1.082_{-0.002}^{+0.002}$ & $5273.48_{-63.38}^{+63.01}$ & $4.29_{-0.08}^{+0.08}$ & $34.16_{-2.38}^{+2.50}$ &  $9.45_{-0.16}^{+0.18}$ & $1.24_{-0.08}^{+0.08}$ & $1.74_{-0.19}^{+0.24}$ & $0.92_{-0.04}^{+0.05}$ &      1620.17 & 1125 &  1.440 \\
 14 & $1.084_{-0.002}^{+0.002}$ & $5017.23_{-63.82}^{+63.11}$ & $4.44_{-0.10}^{+0.11}$ & $36.44_{-2.77}^{+3.01}$ &  $9.26_{-0.18}^{+0.23}$ & $1.23_{-0.08}^{+0.10}$ & $1.57_{-0.19}^{+0.27}$ & $0.91_{-0.04}^{+0.06}$ &      1610.38 & 1154 &  1.395 \\
 15 &                      1.07 &                     5071.51 &                  4.34 &                  32.09 &                   9.48 &                  1.27 &                  1.93 &                  0.96 &      2620.71 & 1230 &  2.131 \\
\bottomrule
\end{tabular}
\end{adjustbox}
\end{table*}


In addition to the previous analysis, we further examined the iron line components around 6.7 keV by incorporating the \texttt{gauss} component, which denotes a Gaussian line profile. Consequently, we tested the model \texttt{const*tbabs*(diskbb+gauss+powerlaw)*gabs}. The \texttt{gauss} model comprises three parameters: line energy ($E_l$), line width ($\Sigma$), and normalization (norm). The differences observed when including or excluding the Gaussian component are presented in Table \ref{tab:3}.

\begin{table}
\begin{center}
\caption{This table depicts the changes in $\chi^2/\nu$ for different non-relativistic spectral models. The equivalent width of the Gaussian iron line is also shown. The $\chi^2/\nu$ change for Spectrum 3 cannot serve as reliable evidence due to the addition of two Gaussian components.}
\label{tab:3}
\begin{adjustbox}{width=220pt}
\begin{tabular}{cccccc}
\toprule
 \multirow{2}{*}{Spec.}& \multirow{2}{*}{$\chi^2$} &  $\chi^2$ & $\chi^2$ &  \multirow{2}{*}{$\nu$} & EW\\
 & &added \texttt{gabs}  & added \texttt{gauss \& gabs} &  & (keV) \\
\midrule
1  & 3275.03(2.792) & 1267.17(1.080) & 1150.08(0.980) & 1173 & 0.023\\
2  & 3660.47(2.935) & 1361.97(1.092) & 1261.21(1.011) & 1247 & 0.030\\
3  & 5147.00(4.121) & 1480.59(1.185) & 1225.24(0.981)* & 1249 & 0.022*\\
4  & 4547.53(3.667) & 1320.37(1.065) & 1244.80(1.004) & 1240 & 0.025\\
5  & 3274.42(2.684) & 1478.89(1.212) & 1420.25(1.164) & 1220 & 0.027\\
6  & 2233.97(1.904) & 1193.74(1.018) & 1148.14(0.979) & 1173 & 0.033\\
7  & 3673.67(2.937) & 1433.54(1.146) & 1369.94(1.095) & 1251 & 0.020\\
\bottomrule
\end{tabular}
\end{adjustbox}
\end{center}
\end{table}

For earlier data sets, we can effectively confine $E_l$ around 6.7 keV, yielding a decrease in $\chi^2/\nu$. However, for Spectrum 3, adding a single \texttt{gauss} model is inadequate because $E_l$ cannot be effectively confined around 6.7 keV. We hypothesize that this is due to a significant broadening of the spectral lines. To counter this, we introduced two \texttt{gauss} models, with the second one successfully replicating the iron line.

\begin{figure}
  \includegraphics[width=0.95\linewidth]{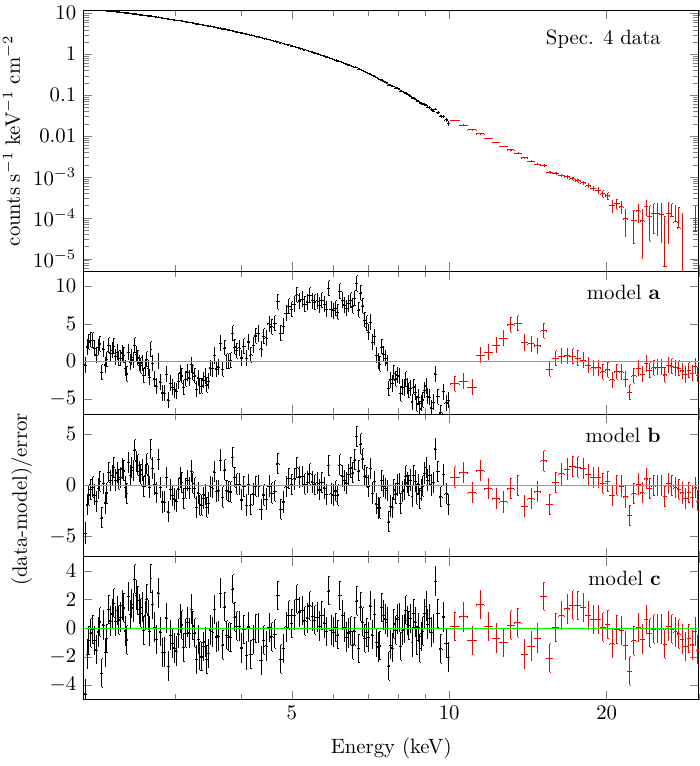}
  \caption{In this figure, we have plotted the data with the folded model of Spec. 4, with residuals displayed using various models: (a) \texttt{const*tbabs*(diskbb+powerlaw)}, (b) \texttt{const*tbabs*(diskbb+powerlaw)*gabs}, and (c) \texttt{const*tbabs*(diskbb+gauss+powerlaw)*gabs}. For plots \textbf{a} and \textbf{b}, we initially ignored the 5 - 8 keV range and then reintroduced this segment. The corresponding $\chi^2/\nu$ values for these models are 3.667, 1.186, and 1.004, respectively.}
  \label{fig:4}
\end{figure}

By utilizing the "eqwidth" command in XSPEC, we were able to estimate the equivalent width of the Gaussian iron line. This information is also tabulated in Table \ref{tab:3}. As the spectrum evolves, the iron line progressively weakens, and $E_l$ cannot be consistently confined around 6.7 keV. Therefore, we focus on those spectra with pronounced iron lines for estimating the spin value. The fit results for different models are shown in Figure \ref{fig:4}.

\subsection{The relativistic spectral model}

Since these spectra exhibit iron line components, we fit them using the relativistic spectral model series \texttt{relxill v2.2}\footnote{\url{http://www.sternwarte.uni-erlangen.de/~dauser/research/relxill/}} \citep{2014MNRAS.444L.100D,2014ApJ...782...76G}. These models are widely utilized to compute the relativistic reflections originating from the innermost regions of the accretion disk around the black hole, playing a significant role in black hole spin measurement \citep{2022SCPMA..6519512F,2022MNRAS.511.3125J}.

However, it is imperative to acknowledge that the \texttt{relxill} series model operates under the assumption that the accretion disk is both optically thick and geometrically thin. This assumption becomes critical given that the luminosity of the source in this study approaches the Eddington limit, which suggests a potential significant transformation of the disk's geometry to a more geometrically thick configuration. Consequently, the application of this model for the precise determination of the spin parameter may not be appropriate. Therefore, it is reasonable to expect that the derived spin values could demonstrate substantial deviations from their true values.

Initially, we employed the fundamental relativistic reflection model \texttt{relxill}, which encompasses a range of parameters. These include the inner and outer disk radius ($R_{\rm{in}}$, $R_{\rm{out}}$), break radius and emissivity index ($R_{\rm{br}}$, $q_1$, $q_2$) used to describe two-part coronal emission, spin parameter ($a_*$), inclination angle ($i$), power-law index ($\Gamma$), ionization of the disk ($\log \xi$), iron abundance ($A_{\rm{Fe}}$), high-energy cutoff ($E_{\rm{cut}}$), and reflection fraction ($R_{\rm{ref}}$). These parameters enable a more comprehensive analysis of the spectral data.

For the fitting process, we continued to hold the $n_{\rm{H}}$ to $4 \times 10^{21} \rm{cm^{-2}}$. The normalization value of \texttt{diskbb} exhibits the property $\rm{norm} = (R_{\rm{in}}/D_{10})^2 \cos \theta$ \citep{1984PASJ...36..741M}, where $R_{\rm{in}}$ denotes the inner disk radius (in km), $\theta$ is the inclination angle of the disk, and $D_{10}$ is the distance to the source (in units of 10 kpc). Considering the binary system's distance as 7.5 kpc and the inclination as 21 degrees \citep{2004ApJ...610..378P}, a norm value of 6000 implies an inner disk radius of 60 km. For a black hole with a mass of 9.5 solar masses and a spin value of 0.7 \citep{2020MNRAS.493.4409D}, the ISCO is roughly 48 km, suggesting that the inner disk radius might be approximately 1.25 times the ISCO radius. Given the closeness of these two values, we assumed that the inner disk radius has reached the ISCO. Hence, we decided to fix $R_{\rm{in}}$ at $-1$. It's important to note that a negative value for $R_{\rm{in}}$ indicates that the unit is in $R_{\rm{ISCO}}$. Additionally, we maintained $q_1 = q_2 = 3$, which is a classic assumption.

Upon applying this model to analyze some of the data, we found that it was not effective in significantly reducing the $\chi^2/\nu$ value. For instance, we obtained $\chi^2/\nu = 2439.88/1173$ for Spectrum 1 and $\chi^2/\nu = 1755.17/1247$ for Spectrum 2. Furthermore, the fitting results were unstable, often leading to aberrant outcomes such as unusually low inclination angles or extreme spin values.

Instead of relying on \texttt{relxill}, we decided to implement the superior model \texttt{relxillCp}. As a result, our comprehensive model evolved into \texttt{const*tbabs*(diskbb+relxillCp)*gabs}. 
The model \texttt{relxillCp} incorporates a crucial parameter, \texttt{$\log N$}, representing the logarithmic value of the density at $R_{\rm{in}}$. Remarkably, this model allows us to directly fit the accretion disk density within a range of $10^{15}$ to $10^{20} \ \rm{cm^{-3}}$. In contrast, the model \texttt{relxill} assigns a fixed density value of $10^{15} \ \rm{cm^{-3}}$. 
However, the fitting results from this analysis suggest a significantly larger accretion disk density, approximately $10^{18} \ \rm{cm^{-3}}$.

We hypothesized that the density parameter significantly affects the $\chi^2/\nu$ value, given the differing densities presented by the two models. 
If we set the density $\log N$ to the same value of $15$ in the   model \texttt{relxillCp}, the spin parameter for both Spectrum 1 and Spectrum 2 would similarly yield a negative value, even reaching as low as -0.998. 
In the study by \citet{2020MNRAS.493.4409D}, they were able to exclude spins below $a_* < 0.5$ at a 99 percent statistical confidence level, effectively ruling out any retrograde geometries. Therefore, utilizing the model \texttt{relxill} for this analysis would be inappropriate due to its limitations.

Additionally, throughout our fitting process, we noticed that most reflection fractions ($R_{\rm{ref}}$) were around 10, but the uncertainty in these values could not be reliably estimated. The values varied considerably and sometimes even reached the boundary, so we froze them at 10.
The reflection fraction is defined as the ratio of the intensity of the primary source irradiating the disk to the intensity directly escaping to infinity, as outlined in \citet{2016A&A...590A..76D}. The intense gravitational field around the compact object tends to deflect photons towards the disk, and values of $R_f > 1$ are expected. Despite these considerations, encountering a reflection fraction as high as 10 remains exceptionally uncommon. This parameter suggests that the current model may not be ideally suited for the spectra under consideration, implying that the results of the fitting process might be subject to bias.

Regarding the electron temperature in the corona ($KT_{\rm{e}}$), we maintained its default value of 60 keV, given that simulating this parameter tends to be unstable. Our decision was based on the goal of ensuring model stability and mitigating the risk of overfitting while preserving crucial elements that describe the system's physical properties.

While the coronal electron temperature can be much higher, particularly in high luminosity cases, we observed minimal changes in the $\chi^2/\nu$ ratio after adjusting the coronal electron temperature to 100 keV or 200 keV. Upon re-fitting, other parameters also showed only slight changes. We believe this is because the spectrum shapes are more significantly influenced by other parameters.

During our fitting process, we observed that the results often varied, leading to different best-fit parameters, such as variations in iron abundance or ionization parameters. However, it is noteworthy that the spin value and inclination parameter remained consistently stable throughout these variations.

\begin{table*}
\caption{This table presents the best-fitting parameters for the model \texttt{const*tbabs*(diskbb+relxillCp)*gabs}. The associated errors, corresponding to a 90\% confidence interval, are also included. We attempted to compute all upper and lower confidence limits using the "error" method in XSPEC.}
\label{tab:4}
\begin{tabular}{lrrrrrr}
\toprule
Spec. &                       2 &                          3 &                           4 &                         5 &                           6 &                         7 \\
\midrule
 \multicolumn{7}{l}{  \texttt{relxillCp} }  \\
$i$ (deg)        &    $33.65_{-4.31}^{+3.80}$ &       $29.11_{-4.43}^{+4.76}$ &        $26.77_{-2.19}^{+3.35}$ &    $31.27_{-5.07}^{+4.48}$   &        $36.61_{-4.82}^{+4.07}$  &      $24.54_{b}^{+4.94}$ \\
$a_*$           &     $0.93^a$ &        $0.91_{-0.21}^{b}$ &        $0.89_{-0.22}^{b}$ &       $0.90^{f}$  &        $0.88^b$ &      $0.75^{a}$ \\
$\Gamma$       &        $3.01^{+0.06}_{-0.15}$ &        $2.86_{-0.05}^{+0.15}$ &         $3.20_{-0.06}^{+0.04}$ &      $3.40_{-0.07}^{b}$ &         $3.38_{-0.17}^{b}$ &      $3.40_{-0.02}^{b}$ \\
$\log \xi$       &   $2.01^{+0.16}_{-0.16}$  &        $4.23_{-0.30}^{+0.29}$ &         $2.28_{-1.16}^{+1.43}$  &          $3.04_{-0.16}^{+0.05}$  &          $3.01_{-0.22}^{+0.04}$ &       $3.01_{-0.11}^{+0.07}$ \\
$\log N$        &       $18.17^{+0.74}_{-2.07}$ &          $16.94_{-0.81}^{+0.14}$ &        $18.06_{-0.04}^{+0.99}$ &          $17.99_{-0.46}^{+0.10}$  &        $17.67^a$ &      $17.83_{-0.28}^{+0.91}$ \\
$A_{\rm{Fe}}$         &        $1.77^{+0.70}_{-0.84}$ &         $3.89_{-1.89}^{+2.03}$ &          $2.77_{-0.79}^{+0.73}$ &    $2.32_{-0.53}^{+0.67}$       &          $3.68^a$ &       $1.48_{-0.84}^{+0.96}$ \\
${R_{\rm ref}}$   &  $10^f$ &  $10^f$ &  $10^f$ &  $10^f$ & $10^f$ &  $10.0^f$ \\ 
$N_{\rm{rel}}$  &       $0.031^a$ &          $0.014_{-0.003}^{+0.013}$  &      $0.020_{-0.010}^{+0.011}$  &    $0.003_{-0.001}^{+0.001}$    &      $0.004^a$ &    $0.003_{-0.001}^{+0.001}$ \\
\midrule
 \multicolumn{7}{l}{  \texttt{diskbb} }  \\
$T_{\rm{in}} $ (keV)        &  $1.150_{-0.003}^{+0.003}$ &     $1.112_{-0.006}^{+0.008}$ &      $1.121_{-0.005}^{+0.004}$ &          $1.111_{-0.006}^{+0.006}$  &      $1.099_{-0.008}^{+0.013}$ &   $1.111_{-0.005}^{+0.007}$ \\
$N_{\rm{bb}}$     &     $6562.26_{-139.13}^{+86.50}$ &   $6443.45_{-88.10}^{+174.58}$ &   $6186.31_{-111.37}^{+165.14}$  &  $6022.46_{-119.55}^{+180.77}$ &  $6071.45_{-249.73}^{+126.09}$ &  $5779.62_{-110.78}^{+89.09}$ \\
\midrule
 \multicolumn{7}{l}{  \texttt{gabs} }  \\
$\rm{Line_E}$ (keV)       &     $9.64_{-0.14}^{+0.13}$ &         $9.80_{-0.23}^{+0.15}$ &         $9.97_{-0.15}^{+0.15}$  &  $9.65_{-0.18}^{+0.17}$      &         $9.57_{-0.42}^{+0.38}$ &        $9.55_{-0.19}^{+0.20}$  \\
$\sigma$ (keV)       &     $1.22_{-0.10}^{+0.11}$ &        $1.82_{-0.25}^{+0.31}$ &         $1.52_{-0.11}^{+0.12}$ &      $1.44_{-0.07}^{+0.13}$   &          $1.61_{-0.22}^{+0.28}$ &       $1.32_{-0.12}^{+0.13}$ \\
Strength    &     $0.90_{-0.10}^{+0.02}$ &         $1.59_{-0.50}^{+0.47}$ &          $1.42_{-0.16}^{+0.17}$  &      $0.97_{-0.14}^{+0.10}$    &         $0.95_{-0.10}^{+0.30}$ &        $0.87_{-0.12}^{+0.14}$ \\

\midrule
 \multicolumn{7}{l}{  \texttt{const} }  \\
C       &     $0.95_{-0.02}^{+0.02}$ &       $0.97_{-0.02}^{+0.02}$ &         $0.97_{-0.02}^{+0.02}$ &      $0.96_{-0.02}^{+0.02}$     &         $0.98_{-0.03}^{+0.03}$  &       $0.93_{-0.02}^{+0.02}$  \\
\midrule 
$\chi^2$ &                    1322.51 &                       1263.42 &                         1238.12 &                  1399.13    &                        1145.10 &                      1325.51 \\
$\nu$        &                     1247 &                        1249 &                         1240 &                       1220 &                         1173 &                       1251\\
$\chi^2/\nu$      &                      1.061 &                         1.012 &                          0.998 &                     1.147       &                          0.976 &                        1.060 \\
\bottomrule
\end{tabular}

    \vspace{5pt}
    \begin{flushleft}
        \footnotesize
        \textsuperscript{a}  Parameter uncertainty can't be estimated. \\
        \textsuperscript{b}  Parameter hits the boundary.
        \\
        \textsuperscript{f}  Frozen parameters.
    \end{flushleft}
    
\end{table*}

From the detailed analysis of the spectra, we present the best-fit results in Table \ref{tab:4}. This analysis suggests that the system likely features a low disk inclination angle, around 30 degrees, and a moderate spin parameter, approximately 0.9. The upper and lower confidence limits were computed using the "error" method in XSPEC.

We then used the "steppar" method in XSPEC to analyze the $\chi^2$ parameter changes for the spin and inclination angle. 
Specifically, the spin parameter was varied over 100 steps, covering a range from 0.2 to 0.98. The result of the $\chi^2$ space exploration is depicted in Figure \ref{fig:5}. 
Similarly, the inclination angle was altered over 100 steps within a range of 20 to 40 degrees, with the corresponding $\chi^2$ space illustrated in Figure \ref{fig:6}. Based on the figures showing statistical variation versus parameter values, we can ascertain that the parameter range for the spin value is approximately 0.9, while for the inclination angle, it is about 25 to 35 degrees.

Figure \ref{fig:5} reveals that the spin parameter values for most spectra exceed 0.8, although certain anomalies are observed. For instance, the minimum spin parameter value for Spectrum 5 is noted to be less than 0.3, even falling into the negative range. To address this anomaly, as detailed in Table \ref{tab:4}, the spin parameter value for Spectrum 5 has been adjusted to 0.9 to represent a more realistic local minimum.
Regarding Figure \ref{fig:6}, with the exception of Spectra 6 and 7, the inclination angles for the remaining spectra demonstrate commendable consistency, converging around 30 degrees. 
This methodical exploration allows for a comprehensive understanding of the parameter's influence on the overall model fit.

\section{DISCUSSION}
\subsection{ Joint fit analysis }

To enhance the precision of our spectral analysis, a joint fitting procedure was applied to Spectra 2 through 7, selected for their prominent iron line features. The choice to persist with the model \texttt{const*tbabs*(diskbb+relxillCp)*gabs} for this collective analysis was informed by the relativistic model's $\chi^2/\nu$ ratio approaching 1, indicating a satisfactory fit.

First, we linked the spin value and inclination angle across all spectra, as these parameters are expected to remain consistent. 
To achieve a more precise fit, we allowed additional parameters to vary. Specifically, we let the hydrogen column density ($n_{\rm{H}}$) vary freely, starting with initial values set to $\rm 0.4 \times 10^{22} cm^{-2}$. 
We also linked $n_{\rm{H}}$ across all spectra, under the assumption that this parameter remains constant over the observation period. Finally, all ME spectra were set to share a common \texttt{const} value to facilitate a more unified and consistent analysis. In this context, we kept the inner disk radius ($R_{\rm{in}}$) fixed at -1. The emissivity indices $q_1$ and $q_2$ were also kept at 3, the reflection fraction was fixed at 10, and the corona temperature was fixed at 60 keV.

\begin{figure}
  \centering
  \includegraphics[width=1\linewidth]{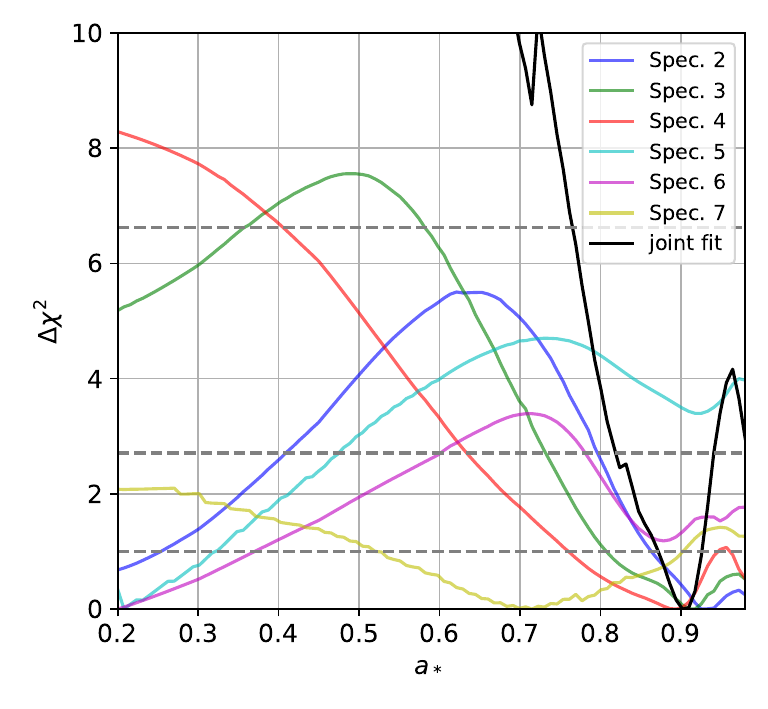}
    \caption{The plot above shows the $\chi^2$ space of the spin parameter, explored with 100 steps ranging from 0.2 to 0.98. The grey dotted lines indicate the 68\%, 90\%, and 99\% confidence levels.}
  \label{fig:5}
\end{figure}

\begin{figure}
  \centering
  \includegraphics[width=1\linewidth]{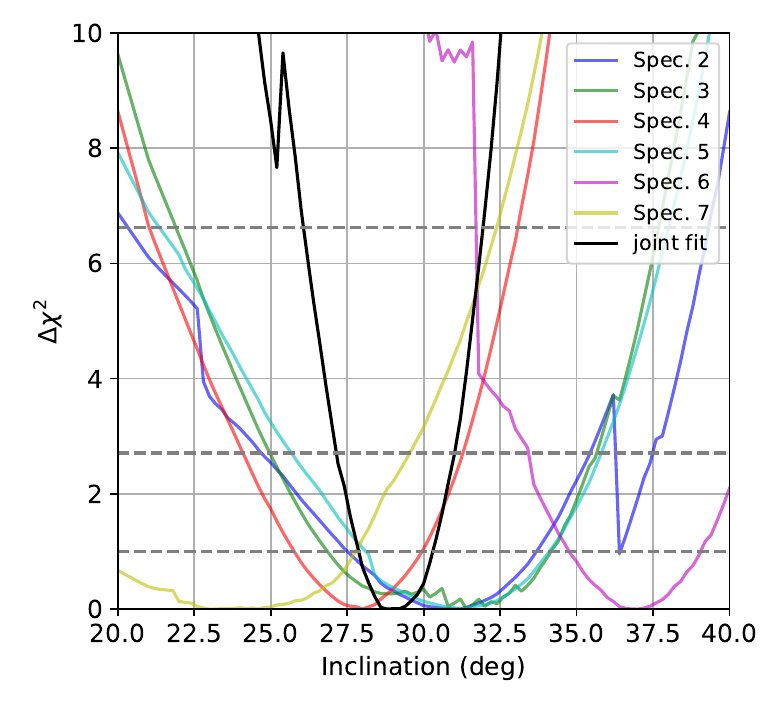}
  \caption{The plot above shows the $\chi^2$ space of the inclination angle parameter, explored with 100 steps ranging from 20 to 40 degrees. The grey dotted lines indicate the 68\%, 90\%, and 99\% confidence levels.}
  \label{fig:6}
\end{figure}

The best-fit values from our analysis are summarized in Table \ref{tab:5}. 
To estimate the uncertainties, we implemented the Markov Chain Monte Carlo (MCMC) method in XSPEC, specifically using the Goodman-Weare algorithm. We employed 90 walkers with a total chain length of 4,000,000 steps, including a 3,000,000-step burn-in phase. The results from all 90 walkers were then combined into a single output file.

The errors in Table \ref{tab:5} are calculated from the chains. We used the "steppar" method to re-examine the spin and inclination angle parameters. The joint fit results from the $\chi^2$ space exploration are depicted in Figures \ref{fig:5} and \ref{fig:6}.

\begin{figure}
  \centering
  \includegraphics[width=1\linewidth]{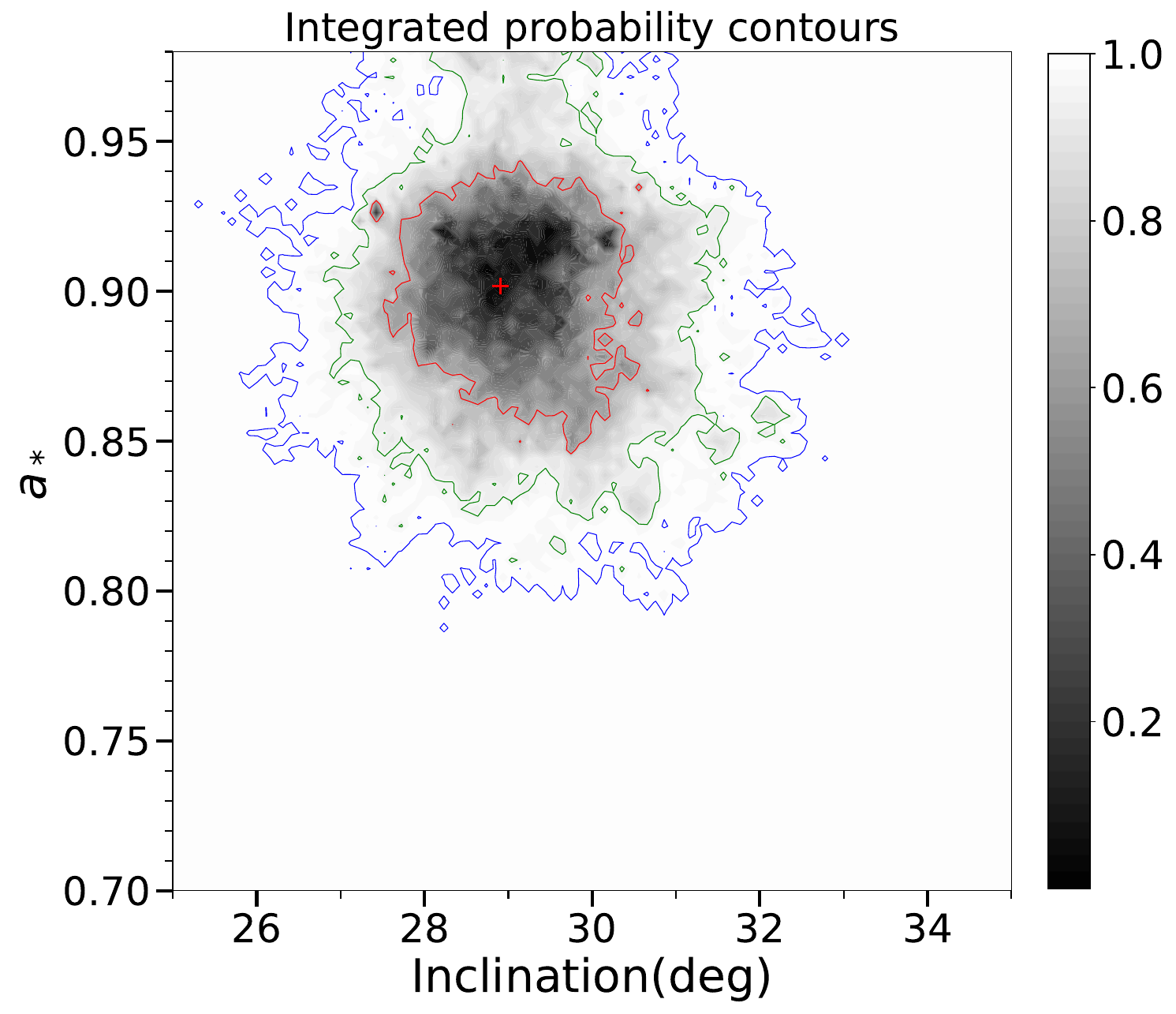}
  \caption{Displayed above is the MCMC probability distribution for the spin parameter $a_*$ and the inclination angle $i$. The range for these two parameters is $0.7$ to  $0.98$ and $25$ to $35$ , respectively, and each is divided into 100 linear bins. The contour lines denote the confidence levels at 99.7\% (3$\sigma$), 95.5\% (2$\sigma$), and 68.3\% ($1\sigma$).}
  \label{fig:7}
\end{figure}

To further investigate the probability distribution of the spin parameter ($a_*$) and the inclination angle ($i$) based on the MCMC results, we calculated the distribution using a grid of 100 bins. We then generated a two-dimensional histogram, depicted in Figure \ref{fig:7}, to visually represent these distributions.

\begin{figure}
  \centering
  \includegraphics[width=1\linewidth]{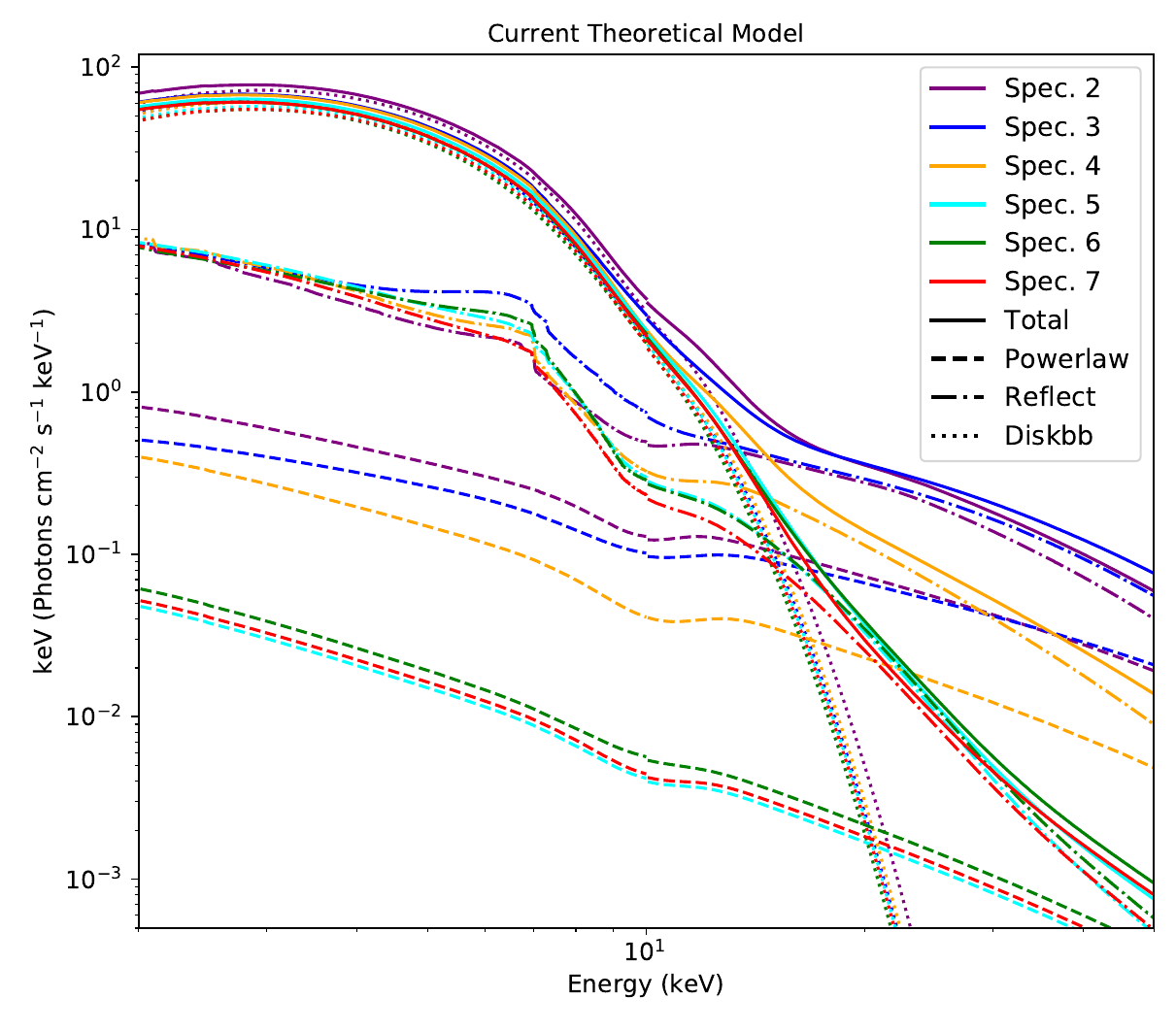}
  \caption{The current theoretical model for \texttt{relxillCp} includes a \texttt{powerlaw} component and a \texttt{reflect} component, represented by dashed and dash-dot lines, respectively. The \texttt{diskbb} component, which dominates the spectrum, is represented by dotted lines.}
  \label{fig:8}
\end{figure}

Figure \ref{fig:8} shows the current theoretical model. In the figures, the lines representing the total, thermal, power-law, and reflection components are depicted as solid, dotted, dashed, and dash-dot lines, respectively.
This indicates that the component \texttt{diskbb} contributes main part to the total emission. Moreover, the  component  \texttt{relxillCp} is used to fit the iron line. 

The spectral energy distribution is predominantly below 10 keV, resulting in an inconspicuous Compton hump within the spectra. Additionally, the component \texttt{diskbb}  does not entirely capture the spectral features, leaving some residuals around 2 keV. To address these residuals, the component \texttt{relxillCp}  inherently adopts a higher reflection fraction, enhancing the fit in the region around 2 keV.  With high R values, the powerlaw component are totally negligible.

\begin{table*}
\begin{center}
\caption{This table shows the best-fit parameters for the model \texttt{const*tbabs*(diskbb+relxillCp)*gabs}, obtained through a joint-fit analysis. The corresponding errors, indicating a 90\% confidence interval, were calculated using the MCMC method. Consistent with Table 4, the emissivity indices were also maintained at 3, the reflection fraction was also fixed at 10, and the corona temperature was similarly held constant at 60 keV.}   

\label{tab:5}
\begin{tabular}{lllllll}
\toprule
Spec. &          2&                   3 &                            4 &       5 &                     6 &                            7 \\
\midrule
 \multicolumn{6}{l}{  \texttt{tbabs} }  \\
$n_{\rm H}$ ($\times 10^{22} \rm{cm}^2$ ) & \multicolumn{6}{c}{$0.400_{-0.012}^{+0.008}$} \\
\midrule
 \multicolumn{6}{l}{  \texttt{relxillCp} }  \\
$i$(deg)     &      \multicolumn{6}{c}{  $28.912_{-1.238}^{+1.824}$   }       \\
$a_*$        &       \multicolumn{6}{c}{  $0.902_{-0.053}^{+0.054}$   }        \\
$R_{\rm{in}}$  &      \multicolumn{6}{c}{  $-1.0^f$  }        \\
$\Gamma$   &       $3.013_{-0.013}^{+0.033}$ &       $2.859_{-0.018}^{+0.028}$ &       $3.213_{-0.025}^{+0.074}$ &       $3.400^{a}$  &       $3.400^{a}$  &       $3.400^{a}$  \\
$\log \xi$ &       $2.015_{-0.137}^{0.094}$ &       $4.211_{-0.202}^{+0.137}$ &       $2.268_{-0.189}^{+0.051}$ &       $3.071_{-0.077}^{+0.039}$  &       $3.047_{-0.164}^{+0.049}$ &       $2.978_{-0.131}^{+0.040}$  \\
$\log N$     &      $18.169_{-0.085}^{+0.125}$ &      $16.874_{-0.134
}^{+0.118}$ &      $18.056_{-0.015}^{+0.045}$ &      $17.945_{-0.105}^{+0.044}$   &      $17.706_{-0.090}^{+0.285}$ &      $17.824_{-0.081}^{+0.167}$\\
$A_{\rm{Fe}}$   &       $1.448_{-0.202}^{+0.129}$ &       $3.655_{-0.691}^{+0.212}$ &       $2.828_{-0.322}^{+0.191}$ &   $2.316_{-0.262}^{+0.115}$ &   $3.582_{-0.375}^{+0.248}$ &   $1.740_{-0.113}^{+0.146}$   \\
$N_{\rm{rel}}$  &       $0.029_{-0.001}^{+0.003}$ &       $0.014_{-0.001}^{+0.001}$ &       $0.020_{-0.001}^{+0.002}$ &       $0.003_{-0.001}^{+0.001}$ &       $0.004_{-0.001}^{+0.001}$  &       $0.004_{-0.001}^{+0.001}$ \\
\midrule
 \multicolumn{5}{l}{  \texttt{diskbb} }  \\
$T_{\rm{in}}$ (keV)  &       $1.151_{-0.002}^{+0.002}$ &       $1.112_{-0.003}^{+0.005}$ &       $1.120_{-0.005}^{+0.003}$ &       $1.111_{-0.002}^{+0.004}$   &       $1.098_{-0.004}^{+0.004}$ &       $1.110_{-0.003}^{+0.001}$\\
$N_{\rm{bb}}$   &  $6558.36_{-94.56}^{+64.00}$ &  $6476.46_{-129.75}^{+100.17}$ &  $6184.84_{-102.90}^{+140.38}$ &  $6032.68_{-128.02}^{+61.19}$ &  $6077.74_{-136.66}^{+113.47}$  &  $5813.37_{-51.73}^{+63.85}$\\
\midrule
 \multicolumn{5}{l}{  \texttt{gabs} }  \\
$\rm{Line_E}$ (keV)    &       $9.676_{-0.081}^{+0.092}$ &       $9.680_{-0.070}^{+0.080}$ &       $10.085_{-0.060}^{+0.114}$ &       $9.647_{-0.086}^{+0.109}$  &       $9.450_{-0.125}^{+0.138}$  &       $9.728_{-0.106}^{+0.093}$\\
$\sigma$ (keV)   &       $1.206_{-0.048}^{+0.080}$ &       $1.728_{-0.109}^{+0.060}$ &       $1.557_{-0.046}^{+0.092}$ &       $1.421_{-0.059}^{+0.088}$ &       $1.556_{-0.116}^{+0.182}$ &       $1.414_{-0.078}^{+0.082}$\\
Strength &       $0.927_{-0.034}^{+0.058}$ &       $1.410_{-0.123}^{+0.079}$ &       $1.558_{-0.062}^{+0.094}$ &       $0.940_{-0.042}^{+0.078}$ &       $0.824_{-0.046}^{+0.053}$  &       $0.993_{-0.068}^{+0.046}$ \\
\midrule
 \multicolumn{7}{l}{  \texttt{Constant} }  \\
C      &      \multicolumn{6}{c}{  $0.959_{-0.009}^{+0.008}$ }  \\
\midrule
$\chi^2$      &     \multicolumn{6}{c}{  $ 7715.62$ }   \\
$\nu$         &     \multicolumn{6}{c}{  $7378 $ }   \\
$\chi^2/\nu$  &     \multicolumn{6}{c}{  $1.046$ }   \\
\bottomrule
\end{tabular}
    \vspace{5pt}
    \begin{flushleft}
        \footnotesize
        \textsuperscript{a}  Parameter uncertainty can't be estimated. \\
        \textsuperscript{b}  Parameter hits the boundary.
        \\
        \textsuperscript{f}  Frozen parameters.
    \end{flushleft}

\end{center}
\end{table*}

\subsection{  Spectrum evolution analysis  }
Indeed, we also attempted to fit other spectra using this model but encountered situations where extreme values were returned. For instance, with Spectrum 1, despite the presence of an iron line, we were unable to obtain a successful estimation for the spin parameter. Interestingly, a negative spin usually yielded a lower $\chi^2/\nu$ value and a seemingly better fit, even resulting in a spin of -0.99. However, such results lack physical validity. Earlier spectra, such as the one with ObsID P0304026001, exhibited similar trends.

We propose that these irregularities arise due to these spectra displaying higher luminosities, reaching and even exceeding the Eddington limit, which is also noted as the brightest black hole transient in over a decade \citep{2023HEAD...2016104C}. Such conditions could induce the emergence of a powerful wind, similar to the phenomenon analyzed in \citet{2023MNRAS.520.4889P}, or the effect of radiation pressure. Both forces have the potential to push the accretion disk outward, causing the inner disk radius to increase. For instance, in the case of Spectrum 1, we set the spin parameter to 0.8 and allowed $R_{\rm{in}}$ to fluctuate freely. This indicated that the radius of the inner disk is approximately 2.2 times the ISCO radius.
With this expanded inner disk radius, it might intuitively seem easier to infer a negative spin value for the black hole. However, a negative spin value lacks physical validity, making it an unsatisfactory explanation.

\begin{figure*}
  \centering
  \includegraphics[width=1\linewidth]{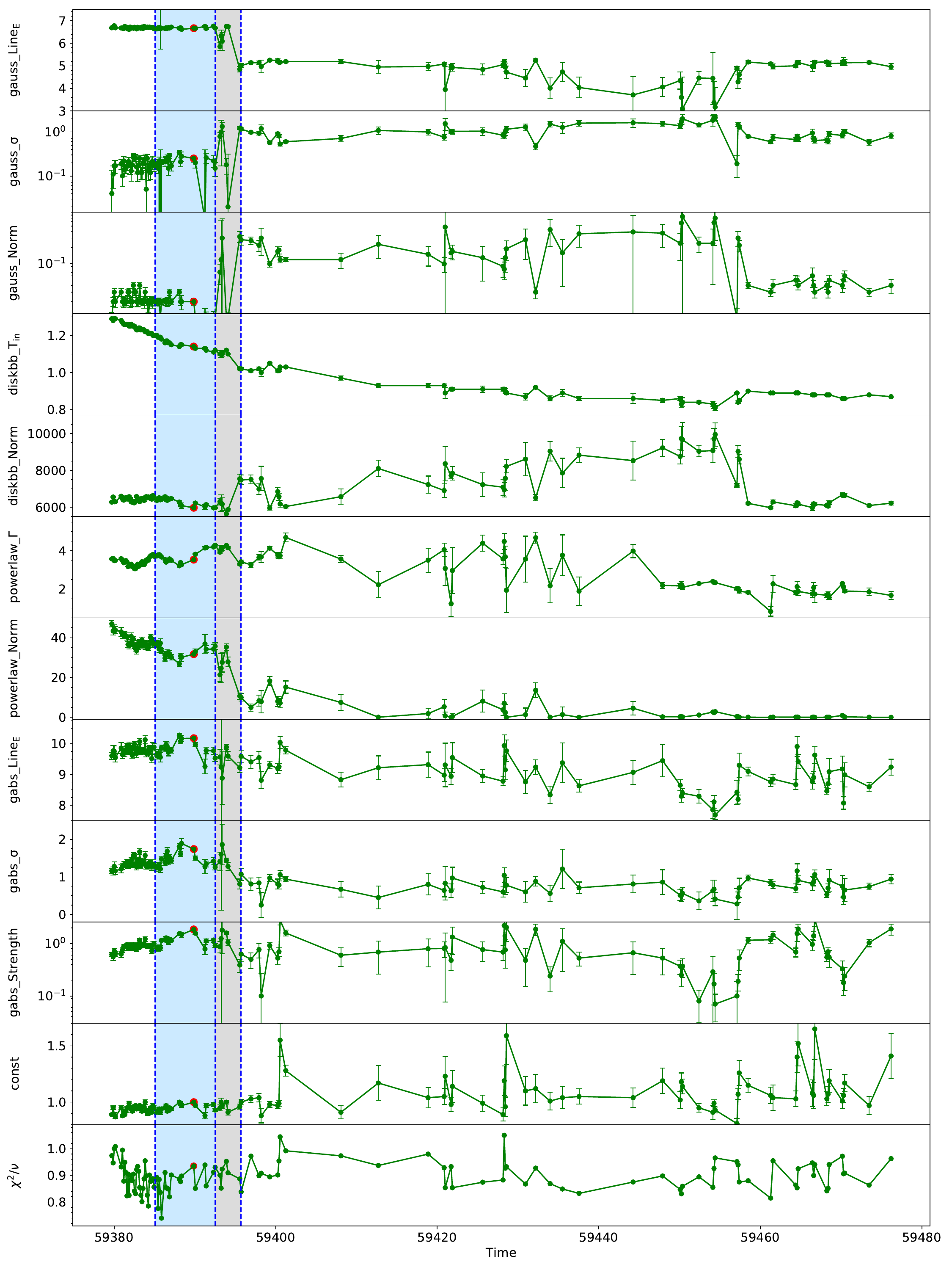}
  \caption{
  Evolution of the spectral parameters for model \texttt{const*tbabs(diskbb+gauss+powerlaw)*gabs}.   }
  \label{fig:9}
\end{figure*}

To elucidate our choice of spectra and explore potential alternatives, we applied the model \texttt{const*tbabs(diskbb+gauss+powerlaw)*gabs} to all spectra, as illustrated in Figure \ref{fig:9}. The shaded blue regions indicate the epochs from Spec. 1 to Spec. 7. This figure reveals that the Gaussian line consistently hovers around 6.67 keV.
In the shaded grey regions, a pronounced line emerges at 5 keV, and the iron line at 6.67 keV begins to fade, suggesting that other spectra may not be suitable for reflection method analysis. Concurrently, the normalization of the powerlaw drops rapidly. We posit that the fading of the iron line is linked to the attenuation of the powerlaw component.
The red dot represents the data from Spec. 3, previously mentioned for requiring two Gaussian components; this is attributed to the emerging line at 5 keV. 

Moreover, other parameter evolutions are evident. The disk temperature diminishes over time, and the disk normalization increases for an extended period before decreasing. Regarding the \texttt{gabs} component, the line center remains between 8-11 keV, with the Strength parameter following a trajectory similar to that reported by \citet{2023MNRAS.520.4889P}.
 
\subsection{  Disk geometry analysis }

From the accretion rate, the structure of the disk can be roughly estimated. The accretion disk can be divided into three regions: the innermost area is the radiation pressure-dominated regime, followed by the radiative cooling-dominated regime, and the outer part is the gas pressure-dominated regime. The physical environments in the two innermost zones were explored by \citet{2006ApJ...648..523W}. In the radiation pressure-dominated regime, under typical circumstances, the vertical thickness, \( H_1 \), is given by:
$$
H_1 = 1.01 \times 10^6 f^{1/2} m \left(\frac{r}{r_{\rm g}}\right)\, \text{cm}
$$
where \( f \) ranges from 0 to 1, and the Schwarzschild radius, \( r_{\rm g} \), is defined as:
$$
r_{\rm g} = 2GM/c^2 = 2.95 \times 10^5 m\, \text{cm}.
$$

In the radiative cooling-dominated regime, the vertical thickness is:
$$
H_2 = 4.43 \times 10^5 \dot{m} m \, \text{cm}.
$$

The boundary between these two regimes is at:
$$
r_{\rm trap} = 0.51\dot{m}
$$

When the luminosity reaches 0.8 times the Eddington luminosity (\(\dot{m}=8\)), the boundary becomes \(r_{\rm trap} = 4.08 r_{\rm g}\). Considering the radiative cooling-dominated regime and employing \(r_{\rm g}\) as the unit of measurement where \(R=kr_{\rm g}\), it follows that:
$$
\frac{H_2}{R} = \frac{H_2}{k r_{\rm g}} = \frac{1.5}{k} \dot{m}.
$$

Considering the luminosity of Spec. 3 - 6 are around 0.8 $\rm L_{Edd}$, by assuming $\dot{m}=8$, we find that $H/R \sim 1$ when $k$ reached 12, this implies that the accretion disk transitions into a thick disk when $R$ is less than $12r_{\rm g}$. 
Based on the assumption that the inner edge of the disk reaches the ISCO, the geometric effect of the disk can no longer be considered negligible. 

In the scenario of a thick accretion disk, \citet{2020MNRAS.491..417R} examined the reflection spectra produced by thick accretion disk and suggested that such configurations could result in an overestimated black hole spin. This potential overestimation arises because the inner edge of a thick disk might extend inward beyond the ISCO. Consequently, the reflection spectrum emerging from the disk’s inner regions could mimic the characteristics of a thinner disk whose inner edge aligns with the ISCO, while also exhibiting a pronounced redshift in some of the emitted photons. 

\begin{figure}
  \centering
  \includegraphics[width=1\linewidth]{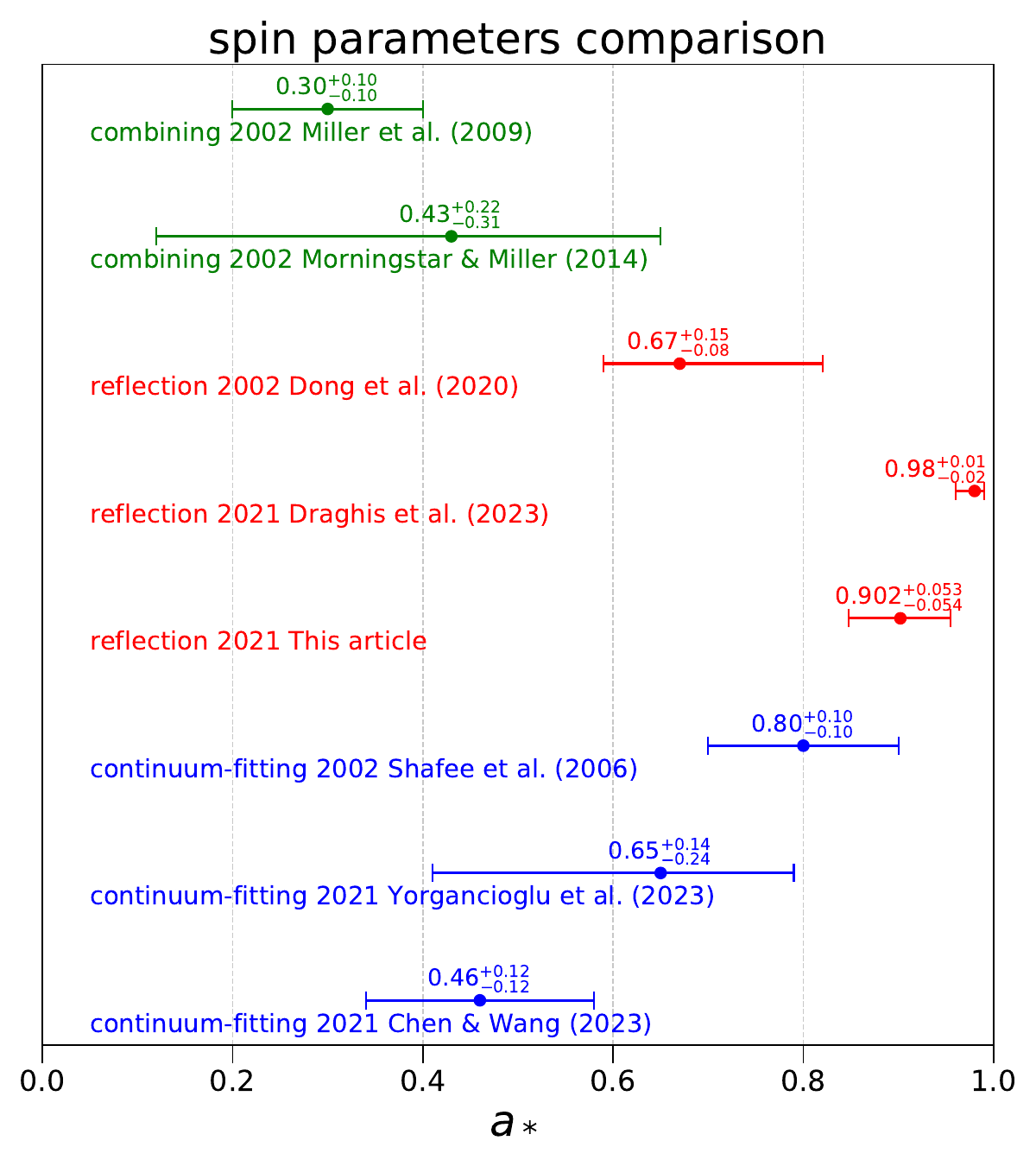}
  \caption{Comparison of spin parameters obtained through different methods. This chart displays the spin parameter values and their associated error margins derived from various methods. The labels "2002" and "2021" indicate the years in which the source experienced bursts, with the resulting data subsequently used for the studies to determine spin values.}
  \label{fig:10}
\end{figure}

Studies of the spin parameters of the source 4U 1543-47 show a wide range of results, suggesting that different methods can lead to varying evaluations. Compared to recent studies, we suggest that in scenarios involving thick disks, the results derived from using the \texttt{relxill} series model are likely overestimated, predominantly due to the influence of high luminosity \citep{2023HEAD...2016104C}. To illustrate these differences, we have included them in Figure \ref{fig:10}, which highlights the discrepancies, particularly in the analyses of the 2002 outburst event.

Similar methods tend to yield consistent results. For instance, both \citet{2014ApJ...793L..33M} and \citet{2009ApJ...697..900M} combined reflection and continuum spectrum analyses and found comparable spin values. 
Further supporting this, the 2021 outburst studies by \citet{2023A&A...677A..79Y} and \citet{2023MNRAS.tmp.2981C} also arrived at similar results using the same methods and HXMT data, enhancing the reliability of their findings.

While many studies report moderate spin parameters, both the analysis by \citet{2023ApJ...946...19D} and our current work, using reflection fitting techniques, have revealed higher spin values. This suggests that for this outburst, reflection fitting may lead to higher spin values.

\section{CONCLUSION}

In our analysis of the 2021 outburst from the black hole X-ray binary system 4U 1543-47, we used data from the \textit{Insight}-HXMT observations. 
Our spectral analysis employed the  reflection model \texttt{relxillCp}  to fit the reflection features observed within the spectra. The findings indicated an inclination angle of the accretion disk to be approximately  $28.91_{-1.24}^{+1.82}$ degrees, and a black hole spin parameter of $0.902_{-0.053}^{+0.054}$, suggesting the existence of a moderately spinning black hole. 
Our study faced challenges with high-luminosity spectra, which yielded extreme and non-physically valid spin values. We hypothesize that luminosities near or beyond the Eddington limit could generate strong winds or radiation pressure effects, causing the inner disk radius to expand and complicating the spin estimation, leading to overestimated results. 
Future research can explore this aspect in greater depth, aiming to improve the robustness of spin measurement methodologies in such scenarios.

Our study showcases the capabilities of the \textit{Insight}-HXMT mission in facilitating precise measurements and in-depth studies of X-ray black hole binaries, thereby contributing to our broader understanding of these cosmic phenomena.

\section*{ACKNOWLEDGEMENTS}
This work is supported by the National Key R\&D Program of China (Grant No. 2023YFA1607902), the National Natural Science Foundation of China (NSFC) (Grant No. 12273058), and science research grants from the China Manned Space Project.
This work made use of the data from the \textit{Insight}-HXMT mission, a project funded by the China National Space Administration (CNSA) and the Chinese Academy of Sciences (CAS). We also thank Mr. Zhao Tong from the National Astronomical Observatories, Chinese Academy of Sciences, for his assistance in polishing the English of this manuscript.

\section*{DATA AVAILABILITY}
The data underlying this article are observed by \textit{Insight}-HXMT, accessed from \url{http://hxmten.ihep.ac.cn/}.



\bibliographystyle{mnras}
\bibliography{bibtex} 

\begin{thebibliography}{}
\makeatletter
\relax
\def\mn@urlcharsother{\let\do\@makeother \do\$\do\&\do\#\do\^\do\_\do\%\do\~}
\def\mn@doi{\begingroup\mn@urlcharsother \@ifnextchar [ {\mn@doi@} {\mn@doi@[]}}
\def\mn@doi@[#1]#2{\def\@tempa{#1}\ifx\@tempa\@empty \href {http://dx.doi.org/#2} {doi:#2}\else \href {http://dx.doi.org/#2} {#1}\fi \endgroup}
\def\mn@eprint#1#2{\mn@eprint@#1:#2::\@nil}
\def\mn@eprint@arXiv#1{\href {http://arxiv.org/abs/#1} {{\tt arXiv:#1}}}
\def\mn@eprint@dblp#1{\href {http://dblp.uni-trier.de/rec/bibtex/#1.xml} {dblp:#1}}
\def\mn@eprint@#1:#2:#3:#4\@nil{\def\@tempa {#1}\def\@tempb {#2}\def\@tempc {#3}\ifx \@tempc \@empty \let \@tempc \@tempb \let \@tempb \@tempa \fi \ifx \@tempb \@empty \def\@tempb {arXiv}\fi \@ifundefined {mn@eprint@\@tempb}{\@tempb:\@tempc}{\expandafter \expandafter \csname mn@eprint@\@tempb\endcsname \expandafter{\@tempc}}}

\bibitem[\protect\citeauthoryear{{Arnaud}}{{Arnaud}}{1996}]{1996ASPC..101...17A}
{Arnaud} K.~A.,  1996, in {Jacoby} G.~H.,  {Barnes} J.,  eds,  Astronomical Society of the Pacific Conference Series Vol. 101, Astronomical Data Analysis Software and Systems V. p.~17

\bibitem[\protect\citeauthoryear{{Bardeen}, {Press}  \& {Teukolsky}}{{Bardeen} et~al.}{1972}]{1972ApJ...178..347B}
{Bardeen} J.~M.,  {Press} W.~H.,   {Teukolsky} S.~A.,  1972, \mn@doi [\apj] {10.1086/151796}, \href {https://ui.adsabs.harvard.edu/abs/1972ApJ...178..347B} {178, 347}

\bibitem[\protect\citeauthoryear{{Cao} et~al.,}{{Cao} et~al.}{2020}]{2020SCPMA..6349504C}
{Cao} X.,  et~al., 2020, \mn@doi [Science China Physics, Mechanics, and Astronomy] {10.1007/s11433-019-1506-1}, \href {https://ui.adsabs.harvard.edu/abs/2020SCPMA..6349504C} {63, 249504}

\bibitem[\protect\citeauthoryear{{Chen} \& {Wang}}{{Chen} \& {Wang}}{2023}]{2023MNRAS.tmp.2981C}
{Chen} J.,  {Wang} W.,  2023, \mn@doi [\mnras] {10.1093/mnras/stad3126}, \href {https://ui.adsabs.harvard.edu/abs/2023MNRAS.tmp.2981C} {}

\bibitem[\protect\citeauthoryear{{Chen} et~al.,}{{Chen} et~al.}{2020}]{2020SCPMA..6349505C}
{Chen} Y.,  et~al., 2020, \mn@doi [Science China Physics, Mechanics, and Astronomy] {10.1007/s11433-019-1469-5}, \href {https://ui.adsabs.harvard.edu/abs/2020SCPMA..6349505C} {63, 249505}

\bibitem[\protect\citeauthoryear{{Connors}, {Garcia}, {Mastroserio}, {Steiner}, {Grefenstette}, {Harrison}  \& {Tomsick}}{{Connors} et~al.}{2023}]{2023HEAD...2016104C}
{Connors} R.,  {Garcia} J.,  {Mastroserio} G.,  {Steiner} J.,  {Grefenstette} B.,  {Harrison} F.,   {Tomsick} J.,  2023, in AAS/High Energy Astrophysics Division. p. 116.104

\bibitem[\protect\citeauthoryear{{Dauser}, {Garcia}, {Parker}, {Fabian}  \& {Wilms}}{{Dauser} et~al.}{2014}]{2014MNRAS.444L.100D}
{Dauser} T.,  {Garcia} J.,  {Parker} M.~L.,  {Fabian} A.~C.,   {Wilms} J.,  2014, \mn@doi [\mnras] {10.1093/mnrasl/slu125}, \href {https://ui.adsabs.harvard.edu/abs/2014MNRAS.444L.100D} {444, L100}

\bibitem[\protect\citeauthoryear{{Dauser}, {Garc{\'\i}a}, {Walton}, {Eikmann}, {Kallman}, {McClintock}  \& {Wilms}}{{Dauser} et~al.}{2016}]{2016A&A...590A..76D}
{Dauser} T.,  {Garc{\'\i}a} J.,  {Walton} D.~J.,  {Eikmann} W.,  {Kallman} T.,  {McClintock} J.,   {Wilms} J.,  2016, \mn@doi [\aap] {10.1051/0004-6361/201628135}, \href {https://ui.adsabs.harvard.edu/abs/2016A&A...590A..76D} {590, A76}

\bibitem[\protect\citeauthoryear{{Dong}, {Garc{\'\i}a}, {Steiner}  \& {Gou}}{{Dong} et~al.}{2020}]{2020MNRAS.493.4409D}
{Dong} Y.,  {Garc{\'\i}a} J.~A.,  {Steiner} J.~F.,   {Gou} L.,  2020, \mn@doi [\mnras] {10.1093/mnras/staa606}, \href {https://ui.adsabs.harvard.edu/abs/2020MNRAS.493.4409D} {493, 4409}

\bibitem[\protect\citeauthoryear{{Draghis}, {Miller}, {Zoghbi}, {Reynolds}, {Costantini}, {Gallo}  \& {Tomsick}}{{Draghis} et~al.}{2023}]{2023ApJ...946...19D}
{Draghis} P.~A.,  {Miller} J.~M.,  {Zoghbi} A.,  {Reynolds} M.,  {Costantini} E.,  {Gallo} L.~C.,   {Tomsick} J.~A.,  2023, \mn@doi [\apj] {10.3847/1538-4357/acafe7}, \href {https://ui.adsabs.harvard.edu/abs/2023ApJ...946...19D} {946, 19}

\bibitem[\protect\citeauthoryear{{Fabian}, {Rees}, {Stella}  \& {White}}{{Fabian} et~al.}{1989}]{1989MNRAS.238..729F}
{Fabian} A.~C.,  {Rees} M.~J.,  {Stella} L.,   {White} N.~E.,  1989, \mn@doi [\mnras] {10.1093/mnras/238.3.729}, \href {https://ui.adsabs.harvard.edu/abs/1989MNRAS.238..729F} {238, 729}

\bibitem[\protect\citeauthoryear{{Feng} et~al.,}{{Feng} et~al.}{2022}]{2022SCPMA..6519512F}
{Feng} Y.,  et~al., 2022, \mn@doi [Science China Physics, Mechanics, and Astronomy] {10.1007/s11433-021-1790-7}, \href {https://ui.adsabs.harvard.edu/abs/2022SCPMA..6519512F} {65, 219512}

\bibitem[\protect\citeauthoryear{{Garc{\'\i}a} et~al.,}{{Garc{\'\i}a} et~al.}{2014}]{2014ApJ...782...76G}
{Garc{\'\i}a} J.,  et~al., 2014, \mn@doi [\apj] {10.1088/0004-637X/782/2/76}, \href {https://ui.adsabs.harvard.edu/abs/2014ApJ...782...76G} {782, 76}

\bibitem[\protect\citeauthoryear{{Harmon}, {Wilson}, {Finger}, {Paciesas}, {Rubin}  \& {Fishman}}{{Harmon} et~al.}{1992}]{1992IAUC.5504....1H}
{Harmon} B.~A.,  {Wilson} R.~B.,  {Finger} M.~H.,  {Paciesas} W.~S.,  {Rubin} B.~C.,   {Fishman} G.~J.,  1992, \iaucirc, \href {https://ui.adsabs.harvard.edu/abs/1992IAUC.5504....1H} {5504, 1}

\bibitem[\protect\citeauthoryear{{Jia} et~al.,}{{Jia} et~al.}{2022}]{2022MNRAS.511.3125J}
{Jia} N.,  et~al., 2022, \mn@doi [\mnras] {10.1093/mnras/stac121}, \href {https://ui.adsabs.harvard.edu/abs/2022MNRAS.511.3125J} {511, 3125}

\bibitem[\protect\citeauthoryear{Kerr}{Kerr}{1963}]{PhysRevLett.11.237}
Kerr R.~P.,  1963, \mn@doi [Phys. Rev. Lett.] {10.1103/PhysRevLett.11.237}, 11, 237

\bibitem[\protect\citeauthoryear{{Kitamoto}, {Miyamoto}, {Tsunemi}, {Makishima}  \& {Nakagawa}}{{Kitamoto} et~al.}{1984}]{1984PASJ...36..799K}
{Kitamoto} S.,  {Miyamoto} S.,  {Tsunemi} H.,  {Makishima} K.,   {Nakagawa} M.,  1984, \pasj, \href {https://ui.adsabs.harvard.edu/abs/1984PASJ...36..799K} {36, 799}

\bibitem[\protect\citeauthoryear{{Liu} et~al.,}{{Liu} et~al.}{2020}]{2020SCPMA..6349503L}
{Liu} C.,  et~al., 2020, \mn@doi [Science China Physics, Mechanics, and Astronomy] {10.1007/s11433-019-1486-x}, \href {https://ui.adsabs.harvard.edu/abs/2020SCPMA..6349503L} {63, 249503}

\bibitem[\protect\citeauthoryear{{Liu}, {Liu}, {Bambi}  \& {Ji}}{{Liu} et~al.}{2022}]{2022MNRAS.512.2082L}
{Liu} Q.,  {Liu} H.,  {Bambi} C.,   {Ji} L.,  2022, \mn@doi [\mnras] {10.1093/mnras/stac616}, \href {https://ui.adsabs.harvard.edu/abs/2022MNRAS.512.2082L} {512, 2082}

\bibitem[\protect\citeauthoryear{{Matilsky}, {Giacconi}, {Gursky}, {Kellogg}  \& {Tananbaum}}{{Matilsky} et~al.}{1972}]{1972ApJ...174L..53M}
{Matilsky} T.~A.,  {Giacconi} R.,  {Gursky} H.,  {Kellogg} E.~M.,   {Tananbaum} H.~D.,  1972, \mn@doi [\apjl] {10.1086/180947}, \href {https://ui.adsabs.harvard.edu/abs/1972ApJ...174L..53M} {174, L53}

\bibitem[\protect\citeauthoryear{{Miller} et~al.,}{{Miller} et~al.}{2002}]{2002ApJ...570L..69M}
{Miller} J.~M.,  et~al., 2002, \mn@doi [\apjl] {10.1086/341099}, \href {https://ui.adsabs.harvard.edu/abs/2002ApJ...570L..69M} {570, L69}

\bibitem[\protect\citeauthoryear{{Miller}, {Reynolds}, {Fabian}, {Miniutti}  \& {Gallo}}{{Miller} et~al.}{2009}]{2009ApJ...697..900M}
{Miller} J.~M.,  {Reynolds} C.~S.,  {Fabian} A.~C.,  {Miniutti} G.,   {Gallo} L.~C.,  2009, \mn@doi [\apj] {10.1088/0004-637X/697/1/900}, \href {https://ui.adsabs.harvard.edu/abs/2009ApJ...697..900M} {697, 900}

\bibitem[\protect\citeauthoryear{{Mitsuda} et~al.,}{{Mitsuda} et~al.}{1984}]{1984PASJ...36..741M}
{Mitsuda} K.,  et~al., 1984, \pasj, \href {https://ui.adsabs.harvard.edu/abs/1984PASJ...36..741M} {36, 741}

\bibitem[\protect\citeauthoryear{{Morningstar} \& {Miller}}{{Morningstar} \& {Miller}}{2014}]{2014ApJ...793L..33M}
{Morningstar} W.~R.,  {Miller} J.~M.,  2014, \mn@doi [\apjl] {10.1088/2041-8205/793/2/L33}, \href {https://ui.adsabs.harvard.edu/abs/2014ApJ...793L..33M} {793, L33}

\bibitem[\protect\citeauthoryear{{Negoro} et~al.,}{{Negoro} et~al.}{2021}]{2021ATel14701....1N}
{Negoro} H.,  et~al., 2021, The Astronomer's Telegram, \href {https://ui.adsabs.harvard.edu/abs/2021ATel14701....1N} {14701, 1}

\bibitem[\protect\citeauthoryear{{Park} et~al.,}{{Park} et~al.}{2004}]{2004ApJ...610..378P}
{Park} S.~Q.,  et~al., 2004, \mn@doi [\apj] {10.1086/421511}, \href {https://ui.adsabs.harvard.edu/abs/2004ApJ...610..378P} {610, 378}

\bibitem[\protect\citeauthoryear{{Prabhakar}, {Mandal}, {Bhuvana}  \& {Nandi}}{{Prabhakar} et~al.}{2023}]{2023MNRAS.520.4889P}
{Prabhakar} G.,  {Mandal} S.,  {Bhuvana} G.~R.,   {Nandi} A.,  2023, \mn@doi [\mnras] {10.1093/mnras/stad080}, \href {https://ui.adsabs.harvard.edu/abs/2023MNRAS.520.4889P} {520, 4889}

\bibitem[\protect\citeauthoryear{{Remillard} \& {McClintock}}{{Remillard} \& {McClintock}}{2006}]{2006ARA&A..44...49R}
{Remillard} R.~A.,  {McClintock} J.~E.,  2006, \mn@doi [\araa] {10.1146/annurev.astro.44.051905.092532}, \href {https://ui.adsabs.harvard.edu/abs/2006ARA&A..44...49R} {44, 49}

\bibitem[\protect\citeauthoryear{{Reynolds}}{{Reynolds}}{2019}]{2019NatAs...3...41R}
{Reynolds} C.~S.,  2019, \mn@doi [Nature Astronomy] {10.1038/s41550-018-0665-z}, \href {https://ui.adsabs.harvard.edu/abs/2019NatAs...3...41R} {3, 41}

\bibitem[\protect\citeauthoryear{{Riaz}, {Ayzenberg}, {Bambi}  \& {Nampalliwar}}{{Riaz} et~al.}{2020}]{2020MNRAS.491..417R}
{Riaz} S.,  {Ayzenberg} D.,  {Bambi} C.,   {Nampalliwar} S.,  2020, \mn@doi [\mnras] {10.1093/mnras/stz3022}, \href {https://ui.adsabs.harvard.edu/abs/2020MNRAS.491..417R} {491, 417}

\bibitem[\protect\citeauthoryear{{Russell}, {Fender}, {Hynes}, {Brocksopp}, {Homan}, {Jonker}  \& {Buxton}}{{Russell} et~al.}{2006}]{2006MNRAS.371.1334R}
{Russell} D.~M.,  {Fender} R.~P.,  {Hynes} R.~I.,  {Brocksopp} C.,  {Homan} J.,  {Jonker} P.~G.,   {Buxton} M.~M.,  2006, \mn@doi [\mnras] {10.1111/j.1365-2966.2006.10756.x}, \href {https://ui.adsabs.harvard.edu/abs/2006MNRAS.371.1334R} {371, 1334}

\bibitem[\protect\citeauthoryear{{S{\'a}nchez-Sierras} et~al.,}{{S{\'a}nchez-Sierras} et~al.}{2023}]{2023A&A...673A.104S}
{S{\'a}nchez-Sierras} J.,  et~al., 2023, \mn@doi [\aap] {10.1051/0004-6361/202245682}, \href {https://ui.adsabs.harvard.edu/abs/2023A&A...673A.104S} {673, A104}

\bibitem[\protect\citeauthoryear{{Shafee}, {McClintock}, {Narayan}, {Davis}, {Li}  \& {Remillard}}{{Shafee} et~al.}{2006}]{2006ApJ...636L.113S}
{Shafee} R.,  {McClintock} J.~E.,  {Narayan} R.,  {Davis} S.~W.,  {Li} L.-X.,   {Remillard} R.~A.,  2006, \mn@doi [\apjl] {10.1086/498938}, \href {https://ui.adsabs.harvard.edu/abs/2006ApJ...636L.113S} {636, L113}

\bibitem[\protect\citeauthoryear{{Watarai}}{{Watarai}}{2006}]{2006ApJ...648..523W}
{Watarai} K.-y.,  2006, \mn@doi [\apj] {10.1086/505854}, \href {https://ui.adsabs.harvard.edu/abs/2006ApJ...648..523W} {648, 523}

\bibitem[\protect\citeauthoryear{{Wilms}, {Allen}  \& {McCray}}{{Wilms} et~al.}{2000}]{2000ApJ...542..914W}
{Wilms} J.,  {Allen} A.,   {McCray} R.,  2000, \mn@doi [\apj] {10.1086/317016}, \href {https://ui.adsabs.harvard.edu/abs/2000ApJ...542..914W} {542, 914}

\bibitem[\protect\citeauthoryear{{Yorgancioglu} et~al.,}{{Yorgancioglu} et~al.}{2023}]{2023A&A...677A..79Y}
{Yorgancioglu} E.~S.,  et~al., 2023, \mn@doi [\aap] {10.1051/0004-6361/202346511}, \href {https://ui.adsabs.harvard.edu/abs/2023A&A...677A..79Y} {677, A79}

\bibitem[\protect\citeauthoryear{{Zhang}, {Cui}  \& {Chen}}{{Zhang} et~al.}{1997}]{1997ApJ...482L.155Z}
{Zhang} S.~N.,  {Cui} W.,   {Chen} W.,  1997, \mn@doi [\apjl] {10.1086/310705}, \href {https://ui.adsabs.harvard.edu/abs/1997ApJ...482L.155Z} {482, L155}

\bibitem[\protect\citeauthoryear{{Zhang} et~al.,}{{Zhang} et~al.}{2020}]{2020SCPMA..6349502Z}
{Zhang} S.-N.,  et~al., 2020, \mn@doi [Science China Physics, Mechanics, and Astronomy] {10.1007/s11433-019-1432-6}, \href {https://ui.adsabs.harvard.edu/abs/2020SCPMA..6349502Z} {63, 249502}

\bibitem[\protect\citeauthoryear{{Zhao} et~al.,}{{Zhao} et~al.}{2020}]{2020JHEAp..27...53Z}
{Zhao} X.-S.,  et~al., 2020, \mn@doi [Journal of High Energy Astrophysics] {10.1016/j.jheap.2020.03.001}, \href {https://ui.adsabs.harvard.edu/abs/2020JHEAp..27...53Z} {27, 53}

\bibitem[\protect\citeauthoryear{{Zhao} et~al.,}{{Zhao} et~al.}{2021}]{2021ApJ...916..108Z}
{Zhao} X.,  et~al., 2021, \mn@doi [\apj] {10.3847/1538-4357/ac07a9}, \href {https://ui.adsabs.harvard.edu/abs/2021ApJ...916..108Z} {916, 108}

\makeatother
\end{thebibliography}





\bsp	
\label{lastpage}
\end{document}